%% file: main.tex
\documentclass{article}
\pdfoutput=1
\usepackage{jinstpub}

\usepackage[binary-units=true,separate-uncertainty=true]{siunitx}
\usepackage{lineno,hyperref}
\usepackage{tikz}
\usepackage{mathtools}
\usepackage{caption}
\usepackage{xspace}
\usepackage{booktabs}
\usepackage{siunitx}
\usepackage{textcomp}
\usepackage[percent]{overpic}
\usepackage{hepnames}
\usepackage[list=true,listformat=simple,margin=10pt]{subcaption}
\usepackage{relsize}

\newcommand{\apsq}{\texorpdfstring{\ensuremath{\smash{\mathrm{Allpix}^2}}}{Allpix\textasciicircum 2}\xspace}
\newcommand{\corry}{\emph{Corryvreckan}\xspace}
\newcommand{\CPP}{C\nolinebreak[4]\hspace{-.05em}\raisebox{.2ex}{\relsize{-1}{\textbf{++}}}\xspace}

\title{\corry: A Modular 4D Track Reconstruction and Analysis Software for Test Beam Data}

\input{authors}

\abstract{
\corry is a versatile, highly configurable software with a modular structure designed to reconstruct and analyse test beam and laboratory data.
It caters to the needs of the test beam community by providing a flexible offline event building facility to combine detectors with different readout schemes, with or without trigger information, and includes the possibility to correlate data from multiple devices based on timestamps.

Hit timing information, available with high precision from an increasing number of detectors, can be used in clustering and tracking to reduce combinatorics.
Several algorithms, including an implementation of Millepede-II, are provided for offline alignment.
A graphical user interface enables direct monitoring of the reconstruction progress and can be employed for quasi-online monitoring during data taking.

This work introduces the \corry framework architecture and user interface, and provides a detailed overview of the event building algorithm.
The reconstruction and analysis capabilities are demonstrated with data recorded at the DESY~II Test Beam Facility using the EUDAQ2 data acquisition framework with an EUDET-type beam telescope, a Timepix3 timing reference, a fine-pitch planar silicon sensor with CLICpix2 readout and the AIDA Trigger Logic Unit.
The individual steps of the reconstruction chain are presented in detail.}

\keywords{Software architectures (event data models, frameworks and databases); Analysis and statistical methods; Data processing methods; Pattern recognition, cluster finding, calibration and fitting methods; Solid state detectors}

\begin{document}
\maketitle
\flushbottom


\section{Introduction}
\label{sec:introduction}
\input{introduction}

\section{Architecture of the Framework}
\label{sec:framework}
\input{framework}

\section{Reconstruction \& Analysis of Test Beam Data}
\label{sec:reco}
\input{reconstruction}

\section{Offline Event Building}
\label{sec:eventbuilding}
\input{eventbuilding}

\section{Reconstruction \& Analysis of Frame-based, Triggered \& Data-Driven Detectors}
\label{sec:data}
\input{example}

\section{Conclusions \& Outlook}
\label{sec:conclusion}
\input{conclusions}

\addcontentsline{toc}{section}{Acknowledgements}
\section*{Acknowledgements}
\label{sec:acknowledgements}
\input{acknowledgements}


\addcontentsline{toc}{section}{References}
\bibliography{references}

\end{document}

%% file: authors.tex
\author[a]{Dominik Dannheim,}
\author[a,1]{Katharina Dort,\note{Also at University of Gie\ss en, Germany.}}
\author[b]{Lennart Huth,}
\author[c]{Daniel Hynds,}
\author[a]{Iraklis Kremastiotis,}
\author[a,2]{Jens Kr\"oger,\note{Also at University of Heidelberg, Germany.}}
\author[a]{Magdalena Munker,}
\author[d]{Florian Pitters,}
\author[b]{Paul Sch\"utze,}
\author[b,*]{Simon Spannagel,\note[*]{Corresponding author.}}
\author[a]{Tomas Vanat,}
\author[a,3]{Morag Williams\note{Now at ESRF, 71 Avenue des Martyrs, 38000 Grenoble, France.}}

\affiliation[a]{CERN, Espl. des Particules 1, 1211 Geneva, Switzerland}
\affiliation[b]{DESY, Notkestra\ss e 85, 22607 Hamburg, Germany}
\affiliation[c]{Nikhef, Science Park 105, 1098 Amsterdam, Netherlands}
\affiliation[d]{HEPHY, Nikolsdorfer G. 18, 1050 Wien, Austria}

\emailAdd{simon.spannagel@desy.de}

%% file: introduction.tex
Test beam measurements using beam telescopes for reference track reconstruction have become important tools for detector R\&D.
The operation of detector prototypes in conditions as close as possible to the final deployment situation enables their testing, qualification, and characterisation in terms of efficiency, spatial resolution, and timing performance.
Reconstruction software frameworks specifically written for test beam experiments, such as EUTelescope~\cite{eutelescope_paper}, Judith~\cite{Judith}, Proteus~\cite{Proteus} or Kepler~\cite{Kepler}, have served the community as tools for many years.
With new generations of silicon detectors emerging and a diversification of the R\&D program conducted at test beam facilities, new requirements on flexibility and interoperability pose a challenge to data reconstruction.
Trigger-based devices, e.g. those designed to operate at the HL-LHC~\cite{hl-lhc}, are operated together with trigger-less detectors developed by the linear collider communities of CLIC~\cite{clic-report} and ILC~\cite{ilc-tdr}, or with fully data-driven multi-purpose detectors such as Timepix3~\cite{timepix3_paper}.

The \corry reconstruction and analysis framework has been designed specifically to address the challenges that come with reconstructing data recorded with such heterogeneous setups while at the same time improving the user experience by lowering the entry barrier for new users to analyse their data.
It is a lightweight and fast framework written in modern \CPP, with a modular design.
The core components deal with the parsing of configuration files, the central event loop, and coordinate transformations while separate modules implement the individual steps of the reconstruction process.
Data are exchanged between individual module instantiations using a central clipboard storage.
Its flexible offline event building algorithm allows the reconstruction chain to be adapted to different analysis requirements and to serve particle detectors with a wide variety of readout schemes.
The flexibility of the framework also enables the analysis of data recorded in stand-alone laboratory experiments with single detectors.

\corry is published as free and open source software under the MIT license, the code can be obtained from the project software repository~\cite{corry-repository}.
A comprehensive user manual has been published~\cite{Corrymanual-pub} and is continuously updated as the framework is extended with new features, the most recent version is available online~\cite{Corryusermanual}.
The framework has greatly profited from the development of the \apsq Generic Pixel Detector Simulation Framework~\cite{allpix-squared}, as both frameworks follow similar philosophies in terms of modularity and configurability, and share parts of their code base.
Furthermore, thanks to a dedicated module in \apsq, \corry is able to directly process simulated detector responses with the same reconstruction chain configured to analyse experimental data.

The framework has already been used in a number of different scenarios, such as the reconstruction of data recorded using the high-rate beam telescope of the CLIC detector \& physics (CLICdp) collaboration at the CERN SPS~\cite{flo_tpx3_timing, soi_paper}, the analysis of test beam data from EUDET-type beam telescopes at the DESY~II Test Beam Facility~\cite{katharina_ichep, jens_instr, magdalena_hstd, morag_iprd, thesis_morag}, the qualification of calorimeter modules with minimum ionising particles~\cite{thesis_thorben}, and the reconstruction of silicon pixel detector simulations and their comparison to data~\cite{allpix-squared}.

This paper describes the architecture and functionality of the \corry framework as released in version 2.0.
Section~\ref{sec:framework} summarises the software structure as well as its user interface.
A detailed description of the reconstruction and analysis chain is presented in Section~\ref{sec:reco}, highlighting features unique to this framework.
Section~\ref{sec:eventbuilding} introduces the offline event building algorithm and provides detailed examples for different setups and configurations.
The full reconstruction and analysis chain is demonstrated in Section~\ref{sec:data} using test beam data recorded with a combination of triggered, frame-based and data-driven devices.
Finally, a summary and an outlook to future developments are given in Section~\ref{sec:conclusion}.

%% file: framework.tex
The \corry framework\footnote{The framework is named after a whirlpool between the islands of Jura and Scarba in the Inner Hebrides.} is designed as a modular software where the algorithms for tasks such as pattern recognition, track reconstruction, and data analysis are implemented as independent modules, loaded as needed at run time.
Key components such as user interaction, the event loop, and parsing of configuration or geometry description are handled centrally by the framework core.
This section provides an overview of this functionality.

\subsection{Core Components of the Framework}

\corry provides a command line interface for interaction with the user.
The main configuration file constitutes the only mandatory argument, containing the reconstruction chain definition.
Individual framework parameters can be supplied or overwritten by passing them to the framework directly on the command line, e.g.\ for parameter scans, or to submit runs to a batch processing system.
The output verbosity of the program as a whole as well as of individual modules can be adjusted via configuration file or command line parameters.

An example for the command line interface of \corry is:
\begin{verbatim}
  corry -c main.config -o parameter=5 -v WARNING
\end{verbatim}
where the main configuration file is specified via the \texttt{-c} option, the \texttt{-o} flag allows the definition of an additional framework parameter, and \texttt{-v} option is used to configure verbosity.
A more detailed description of the interface, including all options, can be found in the user manual.

Three different types of modules are known to \corry.
Global modules are meant to operate on data from all of the detectors available, e.g.\ implementing a track-finding algorithm, and are only instantiated once per run.
Detector modules only operate on the data of an individual detector of the setup, e.g.\ for clustering, and can be limited to processing data only for a subset of the available detectors.
Finally, \emph{Device Under Test (DUT)} modules are only created for detectors that are marked as such.
These are typically analysis modules that calculate observables for the detector under investigation.
It is also possible for developers to restrict the use of individual modules to specific detector types at compile-time, which then omits the creation of the corresponding instances by the module manager.
The separation of modules into these categories simplifies the development of new algorithms as their scope is always well-defined and the processing of data from multiple detectors is inherently abstracted from the algorithm by the framework core.

In contrast to other reconstruction frameworks, \corry does not rely on a specific definition of an event, such as all data related to a single trigger decision, but leaves it to the user to configure, which data constitute one event.
The exact definition of what belongs to this chunk is configured by the user via the event building algorithm described in detail in Section~\ref{sec:eventbuilding}.
Data are passed between the individual modules by means of a clipboard onto which each module can place new data or read and alter existing data.
The central event loop takes care of clearing the clipboard content after processing of an event has been finalised, and before the subsequent event is executed.
A persistent storage space is available on the clipboard to cache data required beyond the scope of a single event.
This storage can be used to preserve data beyond the lifetime of a single event, e.g.\ to accumulate tracks for performing alignment, and is only cleared at the end of the run.

\subsection{Coordinate Systems and Transformations}

Within \corry, the local coordinate system for a detector plane is defined as a right-handed Cartesian coordinate system, with the \textit{x} and \textit{y} axes defining the sensor plane, the \textit{z} axis pointing towards the readout side of the sensor.
Its origin is placed at the centre of the active pixel or strip matrix of the sensor.
The global coordinate system for the full detector setup is defined as a right-handed Cartesian coordinate system as well, where the \textit{z} axis points in the direction of the particle beam.
The orientation of a detector is described by extrinsic active rotations around the geometrical centre of the sensor.
Coordinate transformations between local and global coordinate systems for each detector are provided by the detector class of the framework core.

\subsection{Configuration Files}

\begin{figure}[tbp]
  \centering
  \begin{subfigure}[t]{0.45\linewidth}
        \centering
        \setlength{\fboxsep}{0pt}%
        \fbox{\includegraphics[width=\columnwidth]{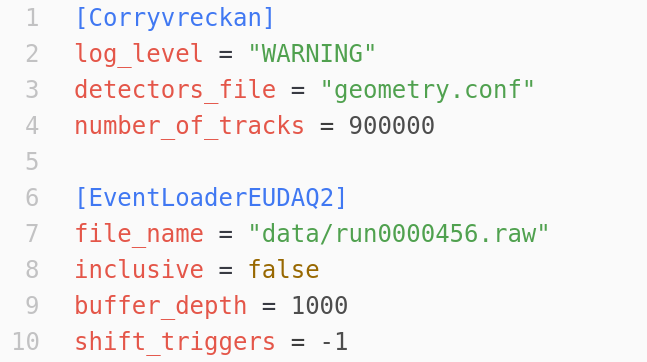}}
        \caption{Example for the main configuration with a global and a module section defining an \emph{EventLoaderEUDAQ2} module and four associated parameters.}
        \label{lst:recoconfig}
    \end{subfigure}\hfill
    \begin{subfigure}[t]{0.45\linewidth}
        \centering
        \setlength{\fboxsep}{0pt}%
        \fbox{\includegraphics[width=\columnwidth]{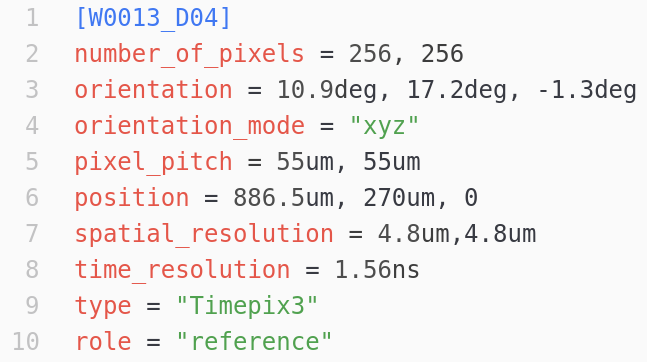}}
        \caption{Example for a detector geometry configuration. The section name \emph{W0013\_D04} corresponds to the detector name, the parameters determine its properties.}
        \label{lst:recogeo}
    \end{subfigure}
    \caption{Excerpts from the two \corry configuration files, the main configuration file~(\subref{lst:recoconfig}) and the geometry description file~(\subref{lst:recogeo}).}
    \label{lst:configs}
\end{figure}

The reconstruction and analysis chain of \corry as well as the detector geometry are configured through text files with an intuitive syntax: a main configuration file and a geometry description file.
These files contain section headers to identify modules, and key/value pairs for the individual configuration parameters.
\corry is capable of interpreting the units provided along with configuration parameters and performs all calculations in a set of internal base units to minimise the need for CPU-intensive conversions and the permanent carrying of units within the code.
If a parameter is provided with units, the value is converted directly into these internal units or otherwise interpreted directly in the base units of the framework.

\subsubsection{Main Configuration File}

The main configuration file contains the structure of the reconstruction chain as well as all relevant module parameters.
Modules are identified by a section header in square brackets, followed by key/value pairs for this specific module as shown in Figure~\ref{lst:recoconfig}.
Here, the \emph{Corryvreckan} section holds global configuration parameters, while the section \emph{EventLoaderEUDAQ2} instantiates a module of the given type.
The four parameters for this particular module are set to a string, a Boolean flag, and integer values.

For each event, modules are executed in the order they are listed in the configuration file.
This is of particular importance for the event building algorithm, which is described in detail in Section~\ref{sec:eventbuilding}.
The configuration parameters available for each module as well as global parameters for the whole framework are listed in the user manual.

\subsubsection{Geometry Configuration File}
\label{sec:geometry}

The properties of all detectors as well as their position and orientation in the global reference frame are specified in the geometry configuration file.
\corry will only process detectors that are present in this geometry, data from other devices are ignored.
Each section of the configuration describes one detector and the section header serves as identifier for the device throughout the reconstruction.
An example for an individual detector is shown in Figure~\ref{lst:recogeo}.
The section contains parameters to describe detector properties such as pixel or strip pitch and matrix size, but also its position and orientation with respect to the global reference frame.

Of special relevance is the \emph{role} parameter that defines how this specific detector behaves in the reconstruction.
It can be configured as DUT so that all DUT-specific modules will be instantiated for this detector; as reference detector against which the alignment is calculated; or as auxiliary device, which is allowed to contribute additional information to the event building process but does not participate in the reconstruction directly.
An example of auxiliary devices are scintillator triggers that can provide additional time stamps.
All other detectors will serve as members of the beam telescope and are used to form reference tracks.

\subsection{Data Model \& Object History}
\label{sec:objects}
\corry provides classes to store different quantities relevant to the reconstruction chain, such as pixel hits, clusters or particle tracks.
These objects inherit from a common \CPP base class that enables storage on the central clipboard of the framework and seamless exchange with other modules and components of the framework as well as writing to and reading from files.

Furthermore, it allows for the storage of inter-object relations via persistent pointers.
Using this, the history of individual objects can be resolved and e.g.\ the original set of pixel hits that lead to the formation of a track can be traced.
This is both possible during processing of the data within \corry and after writing intermediate results to disk.

The separation of the data objects from the algorithmic parts of the framework facilitates the central implementation of track models, which can be used by different, independent track finding modules without reimplementation of the track fitting algorithm.

\subsection{Software Development}
The development of \corry follows best practices for software development by adopting an agile development model, requiring strict format compliance, enforcing a rigorous testing scheme, and by using the C\nolinebreak[4]\hspace{-.05em}\raisebox{.2ex}{\relsize{-1}{\textbf{++}}}14 language standard~\cite{iso-cpp14}.
The code base is well documented; a full class reference is automatically generated from the source code using \texttt{doxygen}~\cite{doxygen} and provided on the website~\cite{corryweb}.
New code contributions to the framework are always reviewed by at least one other developer before merging into the central repository.

Releases of the project follow the semantic versioning scheme, where changes breaking backwards compatibility are only introduced in every new major version, while minor versions add new features and patch versions only provide bug fixes to existing functionality.

Proper functioning of modules and the framework core components as well as compatibility with previous versions of the \corry framework are ensured by a continuous integration, which builds, checks and tests the software on all supported platforms.
This process is integrated with the project software repository and is executed for every new code submission suggested for inclusion.
Here, \emph{checking} refers to directly code-related tasks such as ensuring proper formatting and coding style as well as spell checking, while \emph{testing} implies running complete analyses using reference detector data.
The reference is a centrally hosted, publicly accessible set of data recorded during different test beam campaigns.
They are downloaded automatically by the testing system on demand, and particle tracks are reconstructed and the result compared to a predefined value.
With this data-driven system test, the framework code base can be protected against undesired changes that deteriorate or break existing functionality.
The test suite is extended whenever needed to cover new features or supported detectors.

%% file: reconstruction.tex
The reconstruction chain of \corry is built via the main configuration file by adding modules for each individual task to be performed for the analysis.
This section provides a general overview of a typical configuration of such a reconstruction chain.

\begin{figure}[tbp]
\centering
\fbox{\includegraphics[width=0.9\textwidth]{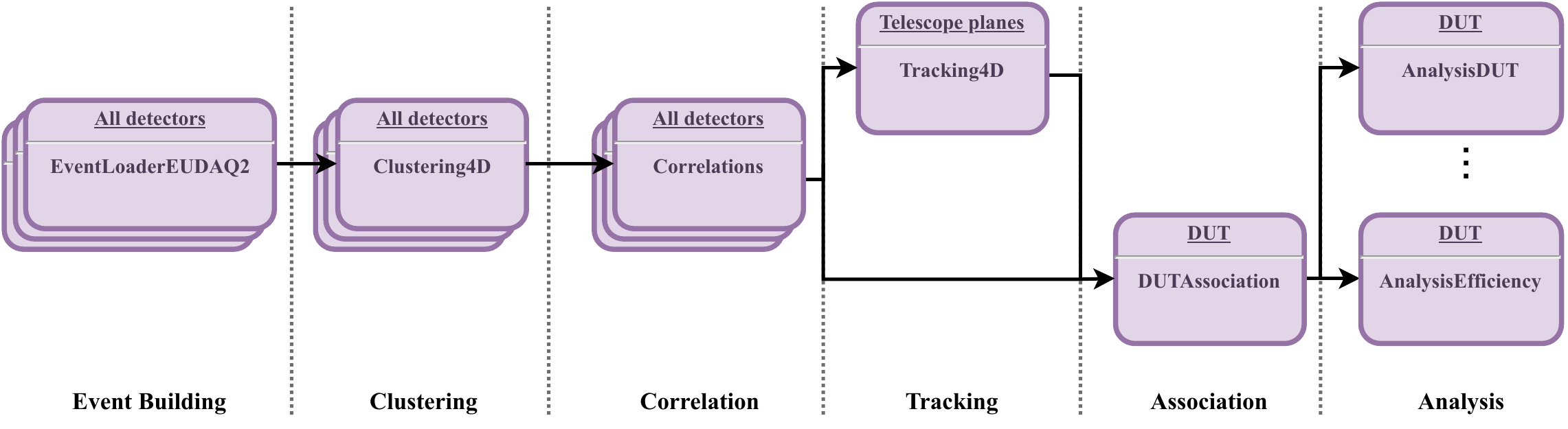}}
\caption{Flow chart of a typical \corry reconstruction chain with a beam telescope setup and a single DUT. Multiple instantiations of a module for different detectors are represented by stacked layers, offsets indicate modules operating on a subset of detectors, and the dots in the \emph{Analysis} stage indicate the possible inclusion of further analysis modules.}
\label{fig:recochain}
\end{figure}

Modules are separated from the core framework and can therefore be flexibly inserted or removed as required to create the desired analysis chain.
A typical example of a reconstruction chain is represented in Figure~\ref{fig:recochain}.
As discussed in Section~\ref{sec:framework}, the order of execution of the modules is defined by their order in the configuration file.
This means they have to be placed at a point in the reconstruction chain where all relevant data are expected to be available.
Each of the different stages of the reconstruction will be briefly discussed in the following.

\subsection{Event Building}
\label{datainput}

The preparation of pixel or strip hit information from all detectors, such as the decoding of raw data or the application of charge or hit-time calibrations, is performed by a group of modules referred to as \emph{EventLoaders}.
\corry comes with a set of detector-specific \emph{EventLoader} modules, which directly decode and process the relevant raw data, but also with the \emph{EventLoaderEUDAQ} and \emph{EventLoaderEUDAQ2} modules, which allow data to be read and used from any detector supported by the EUDAQ1~\cite{eudaq1} or EUDAQ2~\cite{Liu:2019wim} frameworks, respectively.

The first module of the reconstruction chain defines the extent of the currently processed data chunk, defined by means of start and end timestamps or by one or multiple associated trigger numbers.
All subsequent modules have to adhere to this event definition and only provide matching data.

A special role is played by the \emph{FileReader} module that allows intermediate results of a previous reconstruction run or a simulation to be read from a file.
All \corry objects associated with the current event are placed on the clipboard for further treatment.
This allows data to be preprocessed by building events and performing the clustering, while deferring the tracking to a later stage of the data processing.

A detailed description of the event building process with different device combinations is provided in Section~\ref{sec:eventbuilding}.

\subsection{Clustering and Hit Position Interpolation}

\corry offers two different detector modules for clustering, one that performs a closest-neighbour search for all pixel or strip hits on a detector plane, and one that uses pixel hit timestamps as additional criterion for associating pixels with a cluster.
Using timing information drastically reduces combinatorics and allows individual clusters to be recovered accurately in high-rate environments.
Here, strip detectors are treated as detectors with a single row of elongated pixels.

The hit position is calculated from the cluster by means of a charge-weighted centre-of-gravity algorithm if charge information is available.
For binary hit information, the hit position is calculated as the arithmetic mean.

Two modules are provided to perform an $\eta$-correction for non-linear charge sharing~\cite{timepix_tracking} of the cluster centre position in a two-pass approach.
In the first pass, the \emph{EtaCalculation} module is used to record the $\eta$-distribution of the given detector, and to perform a fit of the cumulative $\eta$-function.
The fitting function used can be defined by the user via the configuration file.
The function and the parameters from this fit can then be used in the second pass by the \emph{EtaCorrection} module, which corrects the hit position accordingly.
Only clusters with a width or height of two pixels are considered and corrected by this algorithm.
Since this implementation makes use of the reference tracking information, an independent data set should be used for deriving the correction parameters in order not to bias the results.

\subsection{Track Finding and Fitting}

\corry separates the procedure of finding track candidates from the actual fitting routine of particle tracks.
The track finding algorithms are realised as \emph{Tracking} modules, which have access to the different track models implementing the fit.
In the tracking module, a track model is selected via the configuration and a track candidate is constructed with hits from the relevant detectors.
Then, the fit is performed by the track object and the result reported back to the tracking module.

Different tracking modules are available.
The \emph{Tracking4D} module uses two planes of the telescope to build a first track candidate and subsequently adds more clusters from additional telescope planes if they are within a configurable spatial and time window.
Depending on the detectors and the beam environment, the selection of clusters based on proximity in time as well as space can reduce the number of track candidates to process significantly compared to tracking based only on the spatial correlation of clusters.
The \emph{TrackingMultiplet} module is based on an independent search for particle trajectories, called \emph{tracklets}, in different arms of a beam telescope, usually separated by the DUT.
This approach facilitates an unbiased estimation of the material budget at the kink of the resulting track and is also more efficient at finding particle tracks when a considerable amount of multiple Coulomb scattering is expected from the DUT.

Currently, three different track models are supported and implemented in the respective track objects described in Section~\ref{sec:objects}.
A simple \emph{straight-line track}, described by a reference position and a direction; the \emph{multiplet track} model, described by two tracklets and a single kink between them; and the \emph{General Broken Line (GBL)}~\cite{Blobel2011,gbl} track model, which allows for kinks at every detector and takes into account multiple Coulomb scattering in all detector planes as well as the surrounding air.

\subsection{Association of DUT Clusters with Reference Tracks}

The association between clusters of a DUT and reference tracks built from the telescope planes is implemented as a separate module in \corry.
This facilitates the independent treatment of multiple detectors as DUT at the same time, possibly with different association criteria and configurations.
Multiple clusters from the same detector can be associated with the track, and either all of them or only the associated cluster closest in space can be retrieved from the track for analysis.

The \emph{DUTAssociation} module provides two different ways of associating DUT clusters to the track.
The first compares the cluster centre position to the track position at the DUT and assigns it based on a distance criterion, while the second option makes the association decision based on the closest distance between the track and any of the pixel hits in the cluster.
The latter allows the recovery of large clusters, e.g.\ with contributions from delta rays, where the cluster centre is pulled far away from the track incidence position.

At this step of the reconstruction, it is possible to employ timing information to reduce mismatches between the particle track and the DUT cluster.

\subsection{Prealignment \& Alignment of the Setup}
\label{alignment}

\corry provides the tools necessary for a two-step alignment procedure.
After each of these steps, an updated geometry file in the same format as described in Section~\ref{sec:geometry} is produced, which can be used as input for subsequent alignment steps as well as the final analysis.
Detailed instructions on how to perform alignment, including examples, can be found in the user manual~\cite{Corryusermanual}.

In the first step, a prealignment of all detectors is performed by calculating correlations between all detectors of the setup and the reference plane.
The resulting residuals are then centred around zero by applying shifts to the individual detector positions in the global $x$ and $y$ coordinates.
With this initial prealignment and relatively loose matching criteria it is possible to perform tracking.

These preliminary tracks form the input to the second step of alignment.
Here, different modules are available, which either iteratively refit the individual tracks and minimise the sum of the track $\chi^{2}$ values using Minuit2~\cite{minuit}, or which employ the Millepede-II algorithm~\cite{millepede-proceedings, millepede} to perform a simultaneous fit of all track candidates to determine the alignment corrections.
A dedicated guide to alignment is available as part of the \corry user manual.

\subsection{Analysis of the DUT Performance}

The concept of modularity is also applied to the final analysis of detector performance figures of merit.
Individual modules, such as \emph{AnalysisEfficiency}, \emph{AnalysisSensorEdge} or \emph{AnalysisDUT}, are available to assess for example the efficiency or spatial resolution of each DUT in the reconstruction chain.
These modules can also serve as the basis for a more customised solution for individual analyses.
In addition, an \emph{AnalysisTelescope} module is provided, which facilitates the evaluation of the reference beam telescope performance in terms of tracking efficiency and resolution at the position of the DUT.

Additional modules for specific analysis targets or dedicated to a specific detector prototype can be added to the reconstruction chain as required and users are encouraged to contribute their analysis modules to the code base to make them available to others within the community.

\subsection{Import \& Storage of Reconstruction Data \& Results}

The reconstructed detector clusters and particle trajectories can be stored in the form of \corry data objects as a ROOT TTree~\cite{root} at any point during the reconstruction chain using the \emph{FileWriter} module.
This data format also allows the information to be read back into the framework at a later stage to continue data processing via the \emph{FileReader} module as shown in Section~\ref{datainput}.
Other output modules such as the \emph{TextWriter}, \emph{TreeWriterDUT} or \emph{JSONWriter} only store a selection of the available information, or implement the conversion to different data formats.

In addition, most modules create ROOT histograms from the reconstruction process, which are stored in a central file managed by the framework.
These histograms aid in gauging the quality of the reconstruction process and identifying possible problems, or serve directly as analysis plots for the detector of interest.
Lists of all plots produced by each module can be found in the user manual module descriptions.

In order to facilitate the analysis of detector simulations, the \apsq framework provides a dedicated \emph{CorryvreckanWriter} module, which directly writes files in the format interpreted by the \emph{FileReader} module of \corry.
With this seamless integration, simulation results can be analysed with the same reconstruction chain applied to data.

\subsection{Online Data Quality Monitoring}

\begin{figure}[tbp]
\centering
\includegraphics[width=0.49\textwidth]{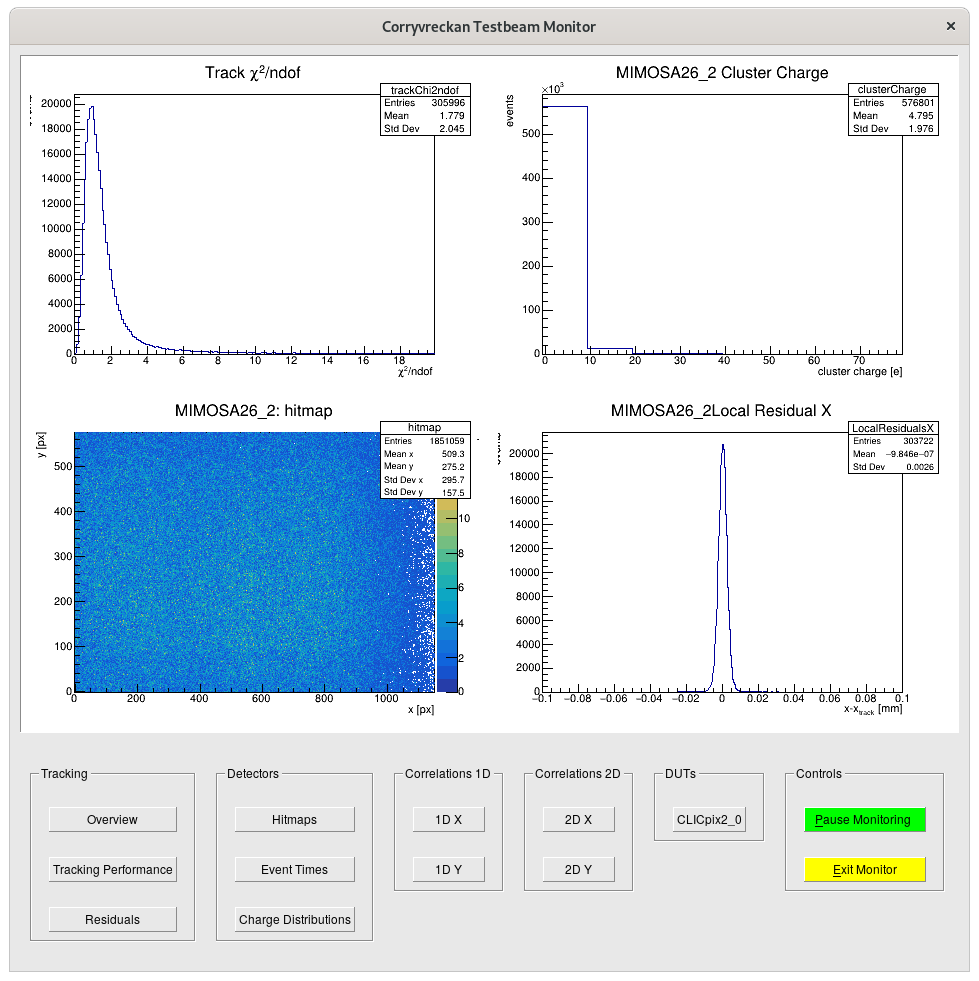}%
\hfill\includegraphics[width=0.49\textwidth]{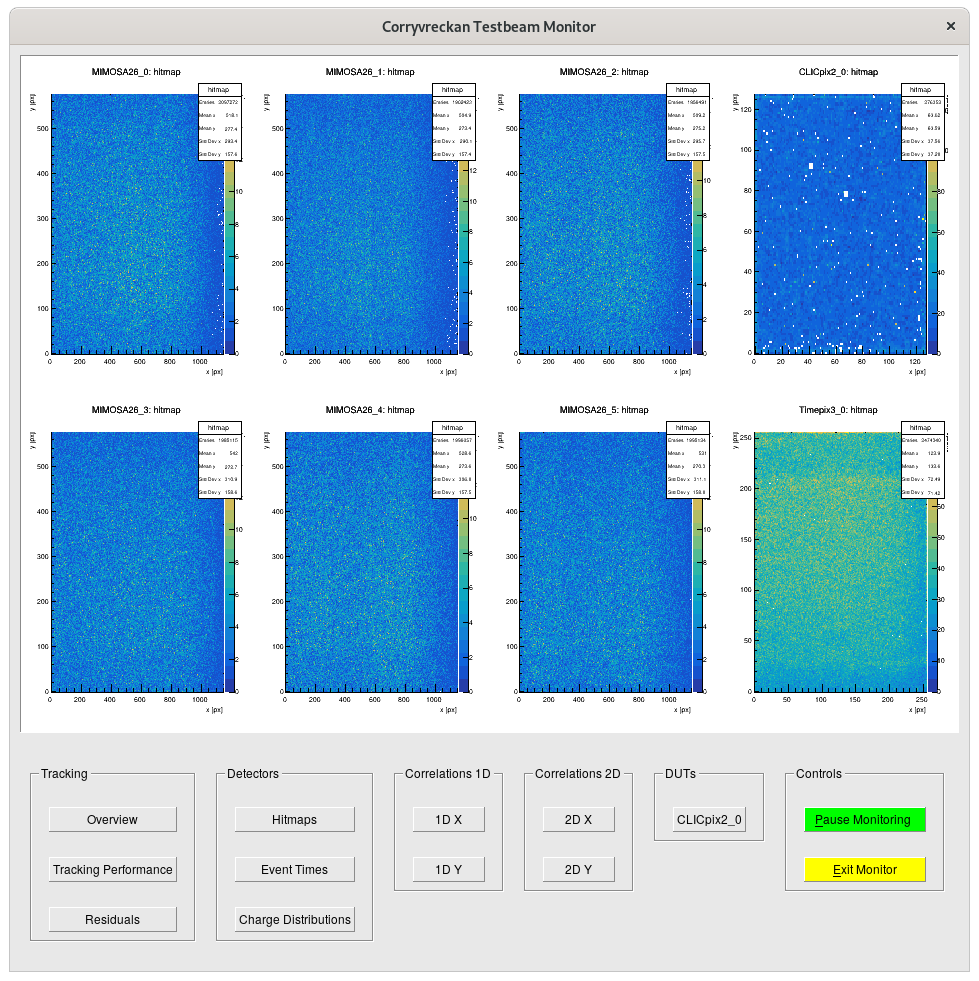}
\caption{The \corry \emph{OnlineMonitor} window showing the quality monitoring overview page~(left) and the detector pixel hit maps~(right). The page selector and control buttons are visible in the lower part of the windows.}
\label{fig:Onlinemonitoringexample}
\end{figure}

The \emph{OnlineMonitor} module turns \corry into a quasi-online data quality monitoring tool.
It provides a graphical user interface, implemented using the ROOT user interface toolkit~\cite{root}, displaying a configurable set of histograms from modules of the current reconstruction run as shown exemplarily in Figure~\ref{fig:Onlinemonitoringexample}.
Multiple pages are available for different stages of the reconstruction such as initial pixel hit maps for all detectors, or correlations and residuals from tracking.
The displayed plots can be selected by the user via the regular \corry configuration files.

The histograms are continually updated as events are processed.
This allows the \emph{OnlineMonitor} to be used for identifying and solving issues in the reconstruction already during data-taking, but also can be used to quickly assess the quality of the data being taken in a test beam environment.

%% file: eventbuilding.tex
One of the key features of \corry is its flexible event building algorithm.
While the reconstruction of data from a heterogeneous set of detectors with similar data and time structure is usually easy to accomplish, test beam experiments often involve the joint operation of prototypes aiming at different experiments and thus require very different readout modes.
This requires an algorithm that enables data to be flexibly collected from all devices in order to successfully reconstruct particle trajectories and measure performance criteria, such as efficiencies.

This section demonstrates the capabilities of the \corry event building algorithm by constructing different events from the same data set for various analysis purposes.
In the following, four types of detectors are distinguished:

\begin{description}
    \item[Trigger Logic Units] (TLUs) provide time information and a corresponding trigger ID for the passage of particles, e.g.\ using the coincidence of different scintillator signals, as well as a common time reference for other detectors in the same setup. An example for such a device is the AIDA-TLU~\cite{Baesso:2019smg}.
    \item[Trigger-based detectors] without individual hit timestamps return blocks of data associated to an externally generated trigger received by the device. Other data, for which no trigger has been received, are discarded. In many cases, the corresponding trigger ID is the only additional information provided alongside the hit data. Examples are detectors developed for the LHC experiments or the Mimosa26 sensors used in the EUDET-type beam telescopes~\cite{Jansen:2016bkd}.
    \item[Frame-based detectors] acquire data in a defined time interval, usually referred to as shutter. The timestamps of the beginning and end of this shutter are often recorded and stored together with the hit data.
    \item[Data-driven detectors] are devices that send their hit data to the DAQ directly after registering the information without requiring an external stimulus for the readout. Therefore, these data usually contain information on hit times using a common clock signal, which is often provided by the TLU for offline synchronisation.
\end{description}

Depending on what information is available, different event building strategies need to be applied as described in the following sections.

\subsection{The Event Building Algorithm}
\label{subsec:eventbuilding}

In \corry, events are formed by first defining the extent of the event in time and by subsequently adding matching data from all detectors that fit into this event.
The \emph{event definition} is performed by the first module in the reconstruction chain, and subsequent modules have to adhere to this definition.
Choosing one or the other detector of the setup to take the role of defining the event enables the selection of different event boundaries, e.g.\ defined by the DUT to only include reference tracks within the active time of this detector.

Usually this task is performed by the \emph{EventLoader} modules introduced in Section~\ref{datainput}.
The module first checks if an event has already been defined and then either adds its data within the event definition, or defines the event before loading data.
Alternatively, special modules can be placed before the first \emph{EventLoader}, allowing the event to be defined independently of a detector.
Examples are the \emph{Metronome} module, which defines successive events of equal length, or the \emph{EventDefinitionM26} module, which correlates information from different devices.

After the event has been defined and all initial data loaded, the full reconstruction chain is executed for the current event, the resulting data is stored and the next event is processed -- defined via the same algorithm.

\subsection{Event Building based on Individual Detectors}

\begin{figure}[tbp]
    \centering
    \fbox{\includegraphics[width=\columnwidth]{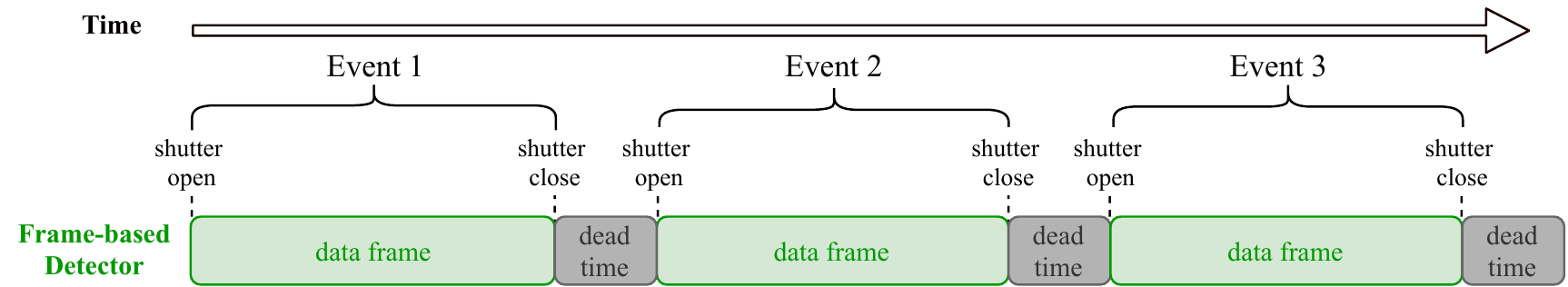}}
    \caption{Schematic of the event definition provided by a device with frame-based readout scheme.}
    \label{fig:framebaseddevice}
\end{figure}

The information available for the event definition depends on the detector used.
For a frame-based readout detector, the time of frame start and end can be used for this definition as shown in Figure \ref{fig:framebaseddevice}.
All subsequent data would have to lie within this defined time frame and would therefore have been recorded during the active time of this detector.
By using this event building definition the efficiency of the frame-based device can be measured correctly despite possibly large dead times outside of the acquisition frame.

\begin{figure}[tbp]
    \fbox{\includegraphics[width=\columnwidth]{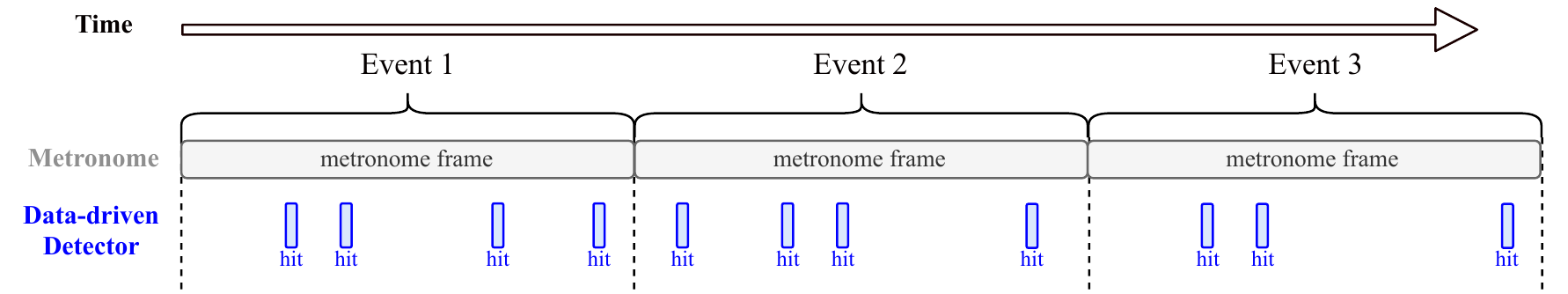}}
    \caption{Schematic of the event definition as provided by the \emph{Metronome} module, generating events of equal length in time. Data from the detector are added within the given time ranges.}
    \label{fig:datadrivendevice}
\end{figure}

In contrast, for a data-driven detector the data stream does not have an inherent structure that would allow a clear separation into events.
Therefore, an arbitrary time structure can be generated using the \emph{Metronome} module as indicated in Figure \ref{fig:datadrivendevice}.
The module defines consecutive events with a configurable length in time such that the data stream is split into time slices of equal length.
If multiple data-driven devices are present, the same definition of an event is applied to all of those devices.
Ambiguities in the attribution of hits near the borders of the event-time windows can be resolved in the event building by applying additional selection criteria taking into account the hit-time resolution of the individual detectors to preserve an unbiased efficiency measurement.

\begin{figure}[tbp]
    \centering
    \fbox{\includegraphics[width=\columnwidth]{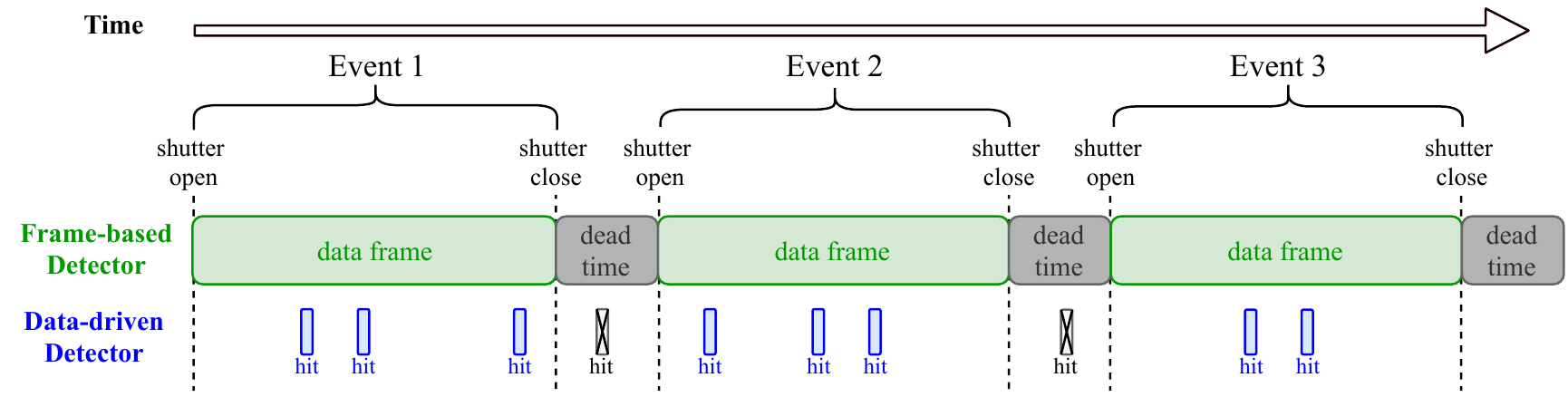}}
    \caption{Schematic of the event definition using the shutter information from a frame-based device and applying the definition on the data stream of the subsequently placed detector.}
    \label{fig:datadrivenandframebased}
\end{figure}

Following this logic it is possible to combine detectors with different readout schemes by ordering their \emph{EventLoader} modules sensibly in the reconstruction chain.
For example, combining a frame-based with a data-driven detector could be performed by defining the event using the shutter information of the frame-based detector to partition the data of the other device as shown in Figure \ref{fig:datadrivenandframebased}.
By employing this event building scheme, hits from the data-driven device that have been recorded outside of the shutter are automatically discarded when loading data.

\begin{figure}[tbp]
    \fbox{\includegraphics[width=\columnwidth]{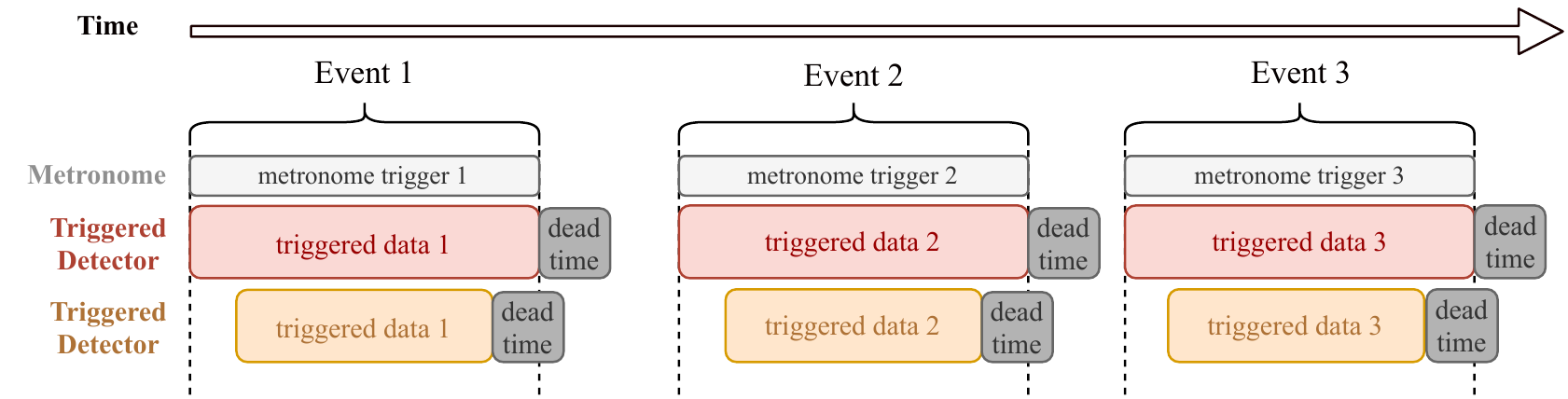}}
    \caption{Schematic of the event definition via event IDs. The \emph{Metronome} defines the event using successive event IDs, and subsequent detectors provide data for matching event IDs.}
    \label{fig:triggereddevice}
\end{figure}

When building events solely from trigger-based devices, events with successive IDs can be defined using the \emph{Metronome} module.
With only trigger-based devices, no time information is necessary and data belonging to the event are identified for each detector by the event ID stored alongside the data as indicated in Figure~\ref{fig:triggereddevice}.
In this case, however, no additional assumption can be made about overlap between data segments and additional measures have to be used to e.g.\ allow for correct efficiency measurements.

\subsection{Event Building Based on a Trigger Logic Unit}

\begin{figure}[tbp]
    \centering
    \fbox{\includegraphics[width=\textwidth]{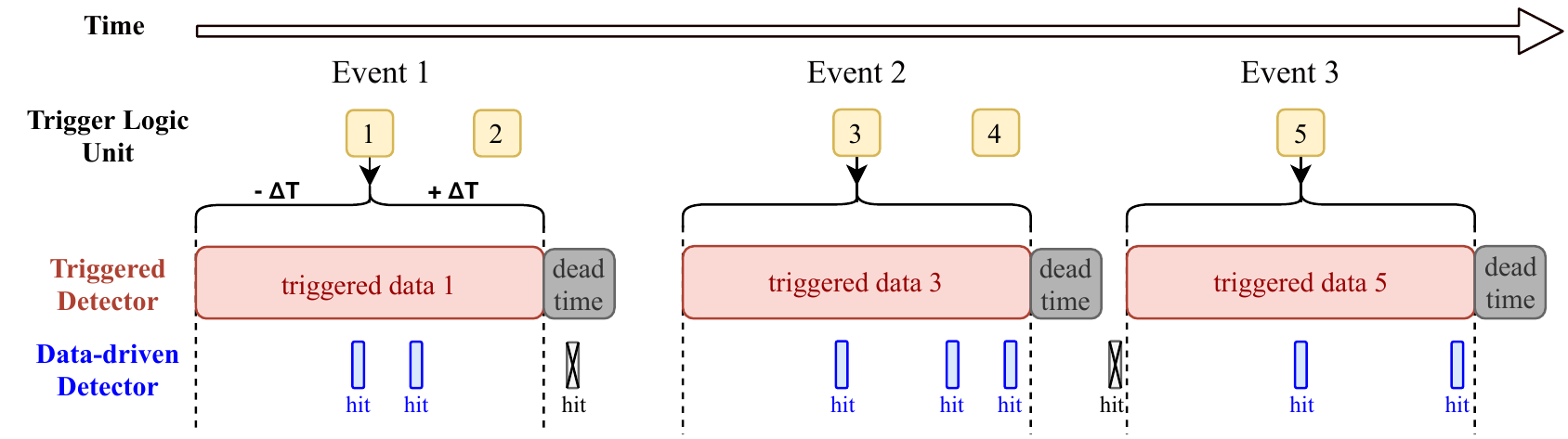}}
    \caption{Schematic of the event definition by a TLU, combining trigger ID and timing information. The event is defined in time by the trigger timestamp and an optional time window indicated by $\pm\Delta T$. Additional triggers during the defined event are added to the definition with their corresponding timestamps.}
    \label{fig:eventbuilding_with_TLU}
\end{figure}

A coherent definition of the event becomes more difficult when detectors with and without time information have to be combined.
In this case, at least one device is required that records both time and trigger information and therefore provides a relation between the other data streams.
This can either be a detector that also provides hit-time information, or a TLU.
The event is then defined both by a trigger ID and the corresponding timestamps.
Optionally, the beginning and end of the event can be defined as a time window around the trigger decision in order to accommodate for the full integration time of the trigger-based device.
An event building scheme taking this into account is illustrated in Figure~\ref{fig:eventbuilding_with_TLU} where the TLU defines the event and subsequent data streams are matched to the event definition either by comparing the trigger number (trigger-based detector) or the beginning and end of the event (data-driven detector).

Similar to the event building around a frame-based device shown in Figure~\ref{fig:datadrivenandframebased}, hits outside the event are discarded.
The IDs of further triggers (such as trigger ID 2 in Figure~\ref{fig:eventbuilding_with_TLU}) arriving during the active window are added to the same event (Event~1 ibid.) instead of defining a new event.
This ensures that all related data from subsequent detectors are collected and avoids re-usage of the same hits for a second event.

\subsection{Event Building with a Frame-based Device with Triggered Shutter}
\label{sec:triggeredshutter}

\begin{figure}[tbp]
    \centering
    \fbox{\includegraphics[width=\textwidth]{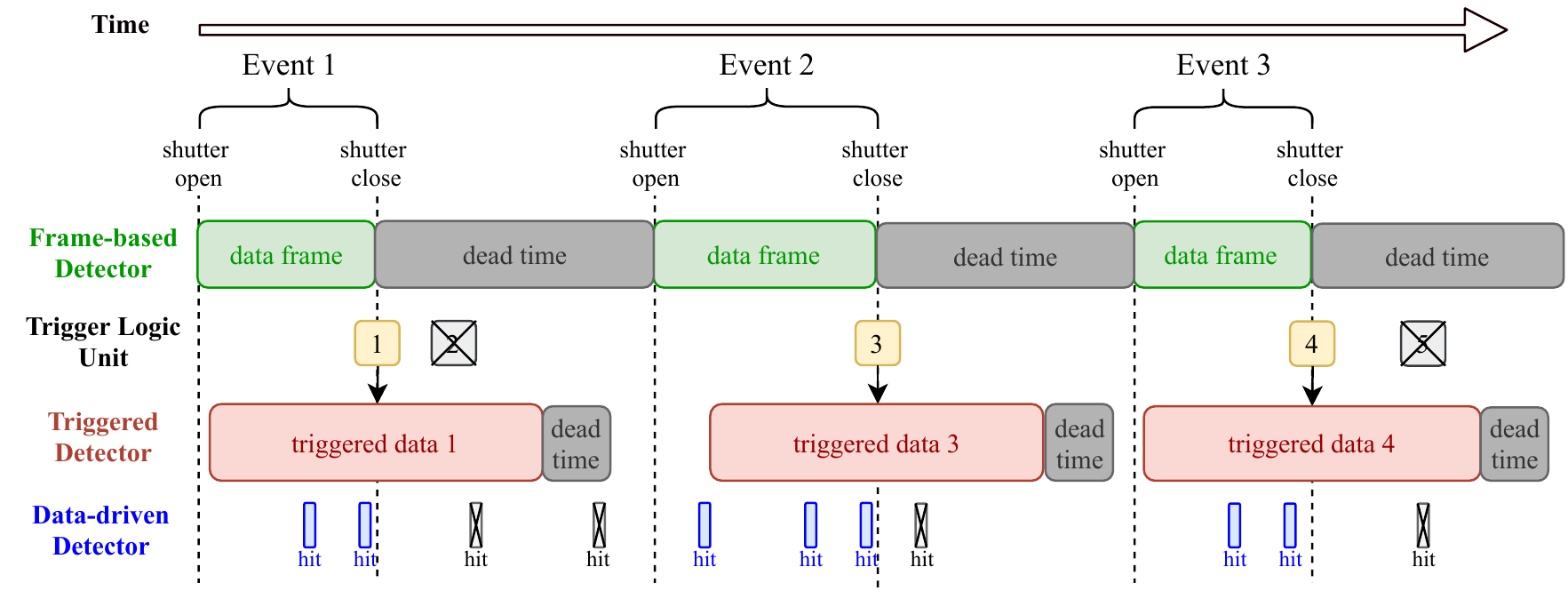}}
    \caption{Schematic of the event building algorithm used for defining events based on a shutter-based device with a triggered shutter-close signal.}
    \label{fig:eventbuilding_shutterbased_plus_TLU}
\end{figure}

Using the event building algorithms described above, also more complex events can be constructed from input data.
Here, a frame-based device whose shutter is controlled by the trigger signal may serve as an example.

The frame-based device has been configured such that its shutter is opened immediately after the dead time of the detector readout, and is kept open until a trigger signal arrives at the detector.
Such a setup can be used, for example, to synchronise the shutter with the arrival time of the particle.
If this detector serves as DUT, only data from within its active shutter period should be taken into account for the reconstruction.
This can be achieved by placing its \emph{EventLoader} module topmost in the reconstruction chain, basing the event definition on its shutter timestamps as depicted in Figure~\ref{fig:eventbuilding_shutterbased_plus_TLU}.
Next, the data stream from the TLU is used in order to add trigger IDs with timestamps within the event boundaries to the definition.
Trigger IDs recorded outside the event window defined by the first module are discarded.
Finally, additional data can be loaded, either based on the begin and end times of the event for data-driven detectors, or the associated trigger IDs for trigger-based detectors.


%% file: example.tex
This section presents an example for a track reconstruction and analysis with \corry, performed on a data set recorded at the DESY~II Test Beam Facility~\cite{desy2_testbeam_facility} with an electron beam of \SI{5.4}{GeV}.
The data acquisition was performed using the EUDAQ2 data acquisition software~\cite{Liu:2019wim}.
In the following, the experimental setup is introduced and the reconstruction chain is discussed in detail.
Finally, a selection of performance parameters of the DUT is analysed.

\subsection{Experimental Setup}

\begin{figure}[tbp]
  \centering
  \fbox{\includegraphics[width=\textwidth]{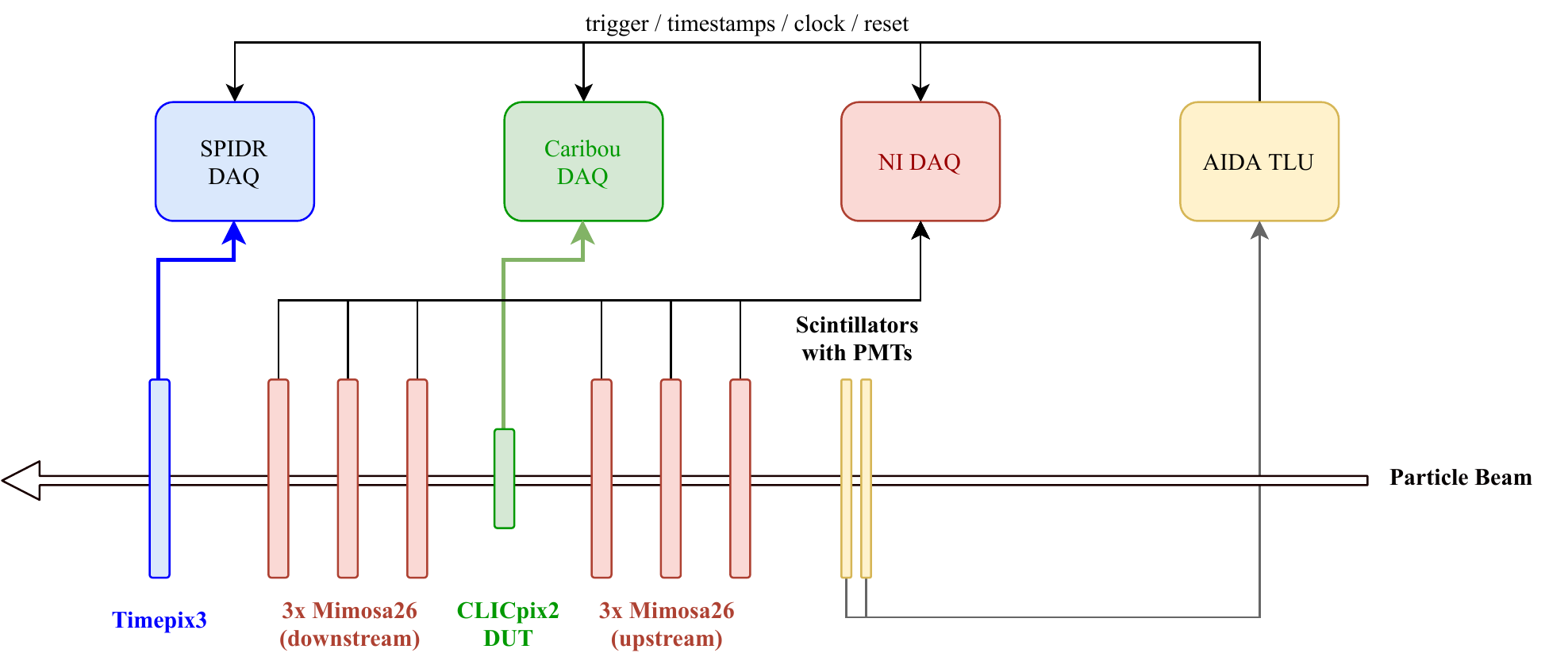}}
  \caption{Schematic drawing of the experimental setup as operated at the DESY~II Test Beam Facility.}
  \label{fig:experimental_setup}
\end{figure}

The experimental setup used to record the data presented in the subsequent sections is shown in Figure~\ref{fig:experimental_setup}.
The following devices have been operated:

\paragraph{The AIDA Trigger Logic Unit}~\cite{Baesso:2019smg}
is used to provide a system-wide time reference as well as the discrimination and coincidence logic to generate trigger signals.
Up to six scintillators equipped with PMTs can be connected to the trigger inputs of the TLU, and an arbitrary coincidence pattern can be configured.
For the data set presented, two scintillators mounted upstream of the beam telescope were operated in coincidence.
In addition to distributing the generated trigger signal to the attached data acquisition systems, the TLU also provides a reference clock as common basis for time measurements.
For this setup, a frequency of \SI{40}{\MHz} has been used.
The clock counter is synchronised between all devices of the setup using a so-called \emph{T0} reset signal distributed at the beginning of a run.

\paragraph{The DATURA telescope}
is a EUDET-type beam telescope~\cite{Jansen:2016bkd} deployed at the beam line 21 of the DESY~II Test Beam Facility.
It consists of six planes of monolithic Mimosa26 sensors with a thickness of \SI{50}{\micro m} each.
The planes are grouped in an upstream and downstream arm, located before and after the DUT.
The sensors are operated in a rolling shutter mode with a period of \SI{115.2}{\micro s}, but only data frames flagged by a trigger signal from the TLU are stored to disk by the NI DAQ system.
A track pointing resolution of about \SI{2}{\micro m} at the DUT can be achieved with this telescope, depending on the geometry of the setup.
Data stored from these devices contain the trigger ID distributed by the TLU.

\paragraph{The Timepix3 timing reference}~\cite{timepix3_paper}
is used for time-tagging of the tracks measured by the DATURA telescope within its data frames.
This allows multiple particle tracks to be disentangled and to unambiguously select those that arrived during the active window of the DUT.
The Timepix3 detector is a hybrid pixel detector with a \SI{100}{\micro m} thick n-in-p planar silicon sensor, a pixel pitch of $\SI{55}{\um} \times \SI{55}{\um}$ with \SI{1.56}{\ns} binned timestamps and a 10-bit charge measurement.
It is operated in a data-driven readout mode providing a data stream of pixel hits with timing information, which is received and processed by the SPIDR DAQ system~\cite{SPIDR}.

The detector was placed downstream of the telescope and is synchronised with the TLU via the common reference clock and the \emph{T0} signal.

\paragraph{The CLICpix2 prototype}~\cite{CLICpix2manual}
is operated as device-under-test.
CLICpix2 is a readout chip designed as a technology demonstrator for the vertex detector of an experiment at the linear lepton collider CLIC.
It features a $128\times128$ pixel matrix with square pixels of \SI{25}{\um} pitch and is operated in a frame-based scheme with external signals controlling the opening and closing of the shutter.
The front-end is capable of simultaneously recording 5-bit Time-over-Threshold (ToT) and 8-bit Time-of-Arrival (ToA) information for each pixel with a time binning of \SI{10}{\ns}.
The assembly used as DUT in the test beam campaign presented consists of a CLICpix2 readout chip bump-bonded to a planar silicon sensor with a thickness of \SI{130}{\um}.
The DUT was operated at a threshold of around 870 electrons, with a bias voltage of \SI{-25}{\V} applied.
For this study, the chip was configured to record data in ToT and counting mode, i.e. replacing the ToA measurement with a particle counter for each pixel.
Therefore, no pixel-level timing information is available for this device.
However, the timestamps of the shutter opening and closing are recorded using the reference clock and \emph{T0} signal from the TLU, allowing the determination of the active window of the detector.
CLICpix2 is operated and read out by the Caribou DAQ system~\cite{caribou}.

\subsection{Reconstruction Chain}

In this section, the reconstruction chain for the test beam data is described in detail following the flow chart presented in Figure~\ref{fig:recochain}, from data decoding and event building to tracking and alignment.

\paragraph{Event building and raw data processing}

\begin{figure}[tbp]
  \centering
  \begin{subfigure}[t]{0.535\linewidth}
    \begin{overpic}[width=\textwidth]{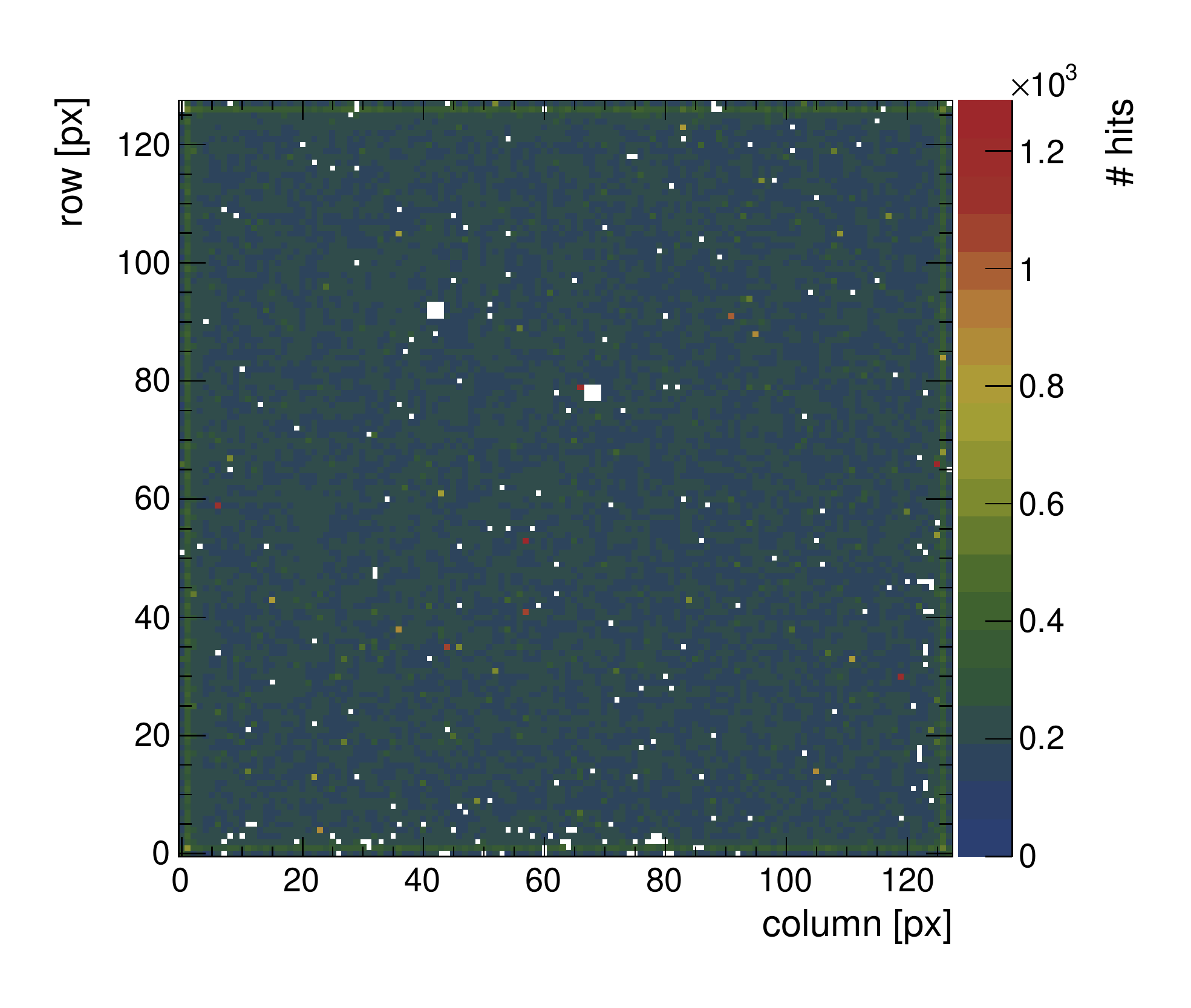}
      \put(20,68){\color{white}CLICdp}
    \end{overpic}
    \caption{Pixel hit map of the CLICpix2 prototype. Masked pixels appear white and pixels with increased occupancy in yellow.}
    \label{fig:hitmap}
  \end{subfigure}%
  \begin{subfigure}[t]{0.465\linewidth}
    \begin{overpic}[width=\textwidth]{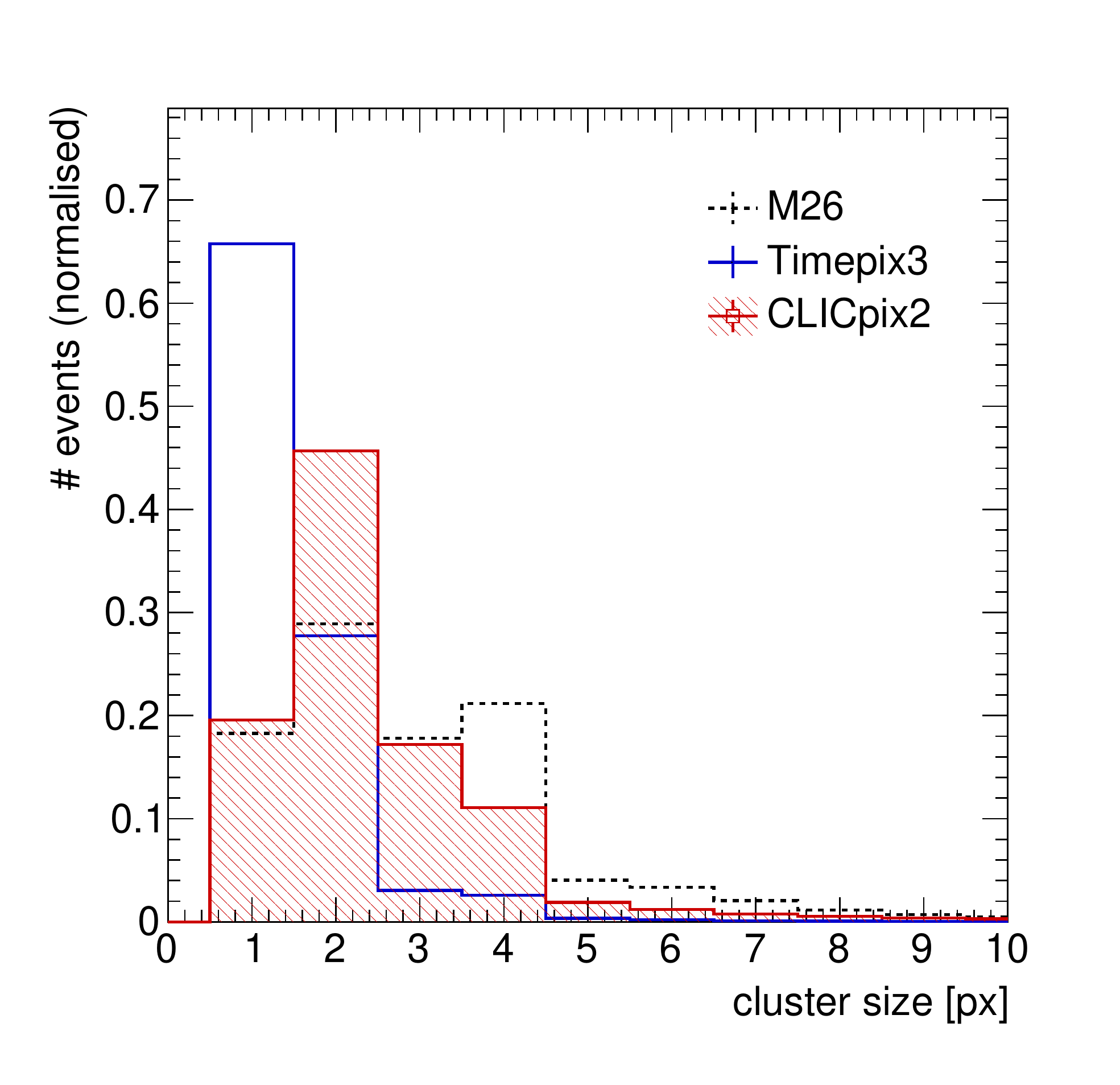}
      \put(20,78){CLICdp}
    \end{overpic}
    \caption{Cluster size distributions: Mimosa26 (M26, dashed black line), Timepix3 (solid blue line) and CLICpix2 (red hatched area).}
    \label{fig:clustersize}
  \end{subfigure}%
  \caption{Pixel hit map of CLICpix2~(\subref{fig:hitmap}) and comparison of the cluster size distribution for the different detectors~(\subref{fig:clustersize}).}
  \label{fig:eventbuilding}
\end{figure}

The event building is based on the shutter opening and closing time of the CLICpix2 prototype detector.
Since all detectors have been operated using the EUDAQ2 DAQ framework, the \emph{EventLoaderEUDAQ2} module is used to read and process the raw data.
It uses the converter plugins implemented directly in the DAQ framework in order to decode the raw detector data to pixel hit information with column, row, charge, and time measurement where applicable.

These decoded data are placed on the \corry clipboard in the order of appearance of the respective \emph{EventLoader} modules in the \corry main configuration file, i.e.\ CLICpix2, TLU, Mimosa26, and Timepix3.
Here, it is important that TLU data is processed before Mimosa26 data because it adds the relevant trigger IDs recorded between shutter open and close signals of the CLICpix2 detector to the event.
These trigger IDs are then used to assign the appropriate Mimosa26 frames to the data block of the event.
The TLU trigger timestamp is assigned to pixel hits received from Mimosa26 sensors.

Calibration of the pixel charge and timestamps is taken care of by the corresponding EUDAQ2 converter plugins, the relevant calibration data can be supplied either through the \corry geometry description or as configuration parameter of the event loader module.
Pixels can be masked offline by providing a mask file to the framework, and hits from masked pixels are filtered out before being stored on the clipboard and therefore appear empty in hit maps.
Mask files can either be populated manually or by using the \emph{MaskCreator} module, which masks pixels based on their firing frequency with respect to either the total chip frequency or a local density estimate.
A hit map for the CLICpix2 DUT is shown in Figure~\ref{fig:hitmap}, which shows that not all pixels with increased occupancy have been filtered out.
The level of acceptable noise hits and therefore the selected masking criterion depends on the individual application situation.

\paragraph{Clustering}
The clustering of pixel hits stemming from the same particle is performed using the \emph{Clustering4D} module, which takes into account the time information assigned to the individual pixel hits.
Adjacent pixel hits are grouped into clusters if their timestamps are within a time window defined in the \corry configuration for each detector individually, taking into account effects such as time walk.

A comparison of the cluster size distributions of the different detectors is shown in Figure~\ref{fig:clustersize}.
While Timepix3 predominantly generates single-pixel clusters owing to its comparatively large pixel pitch, the thin sensor, and the perpendicular particle incidence, Mimosa26 and CLICpix2 produce significantly larger clusters.
In Mimosa26 the reason for this is the signal collection relying on diffusion processes, while CLICpix2 has an increased charge sharing between pixels due to the pixel pitch of \SI{25}{\um} in combination with the sensor thickness of \SI{130}{\um}.

The \emph{EtaCorrection} module is used in order to perform an $\eta$-correction for non-linear charge sharing of the cluster position for the CLICpix2 DUT.

\paragraph{Correlations}

\begin{figure}[tbp]
  \centering
  \begin{subfigure}[t]{0.5\linewidth}
    \begin{overpic}[width=\textwidth]{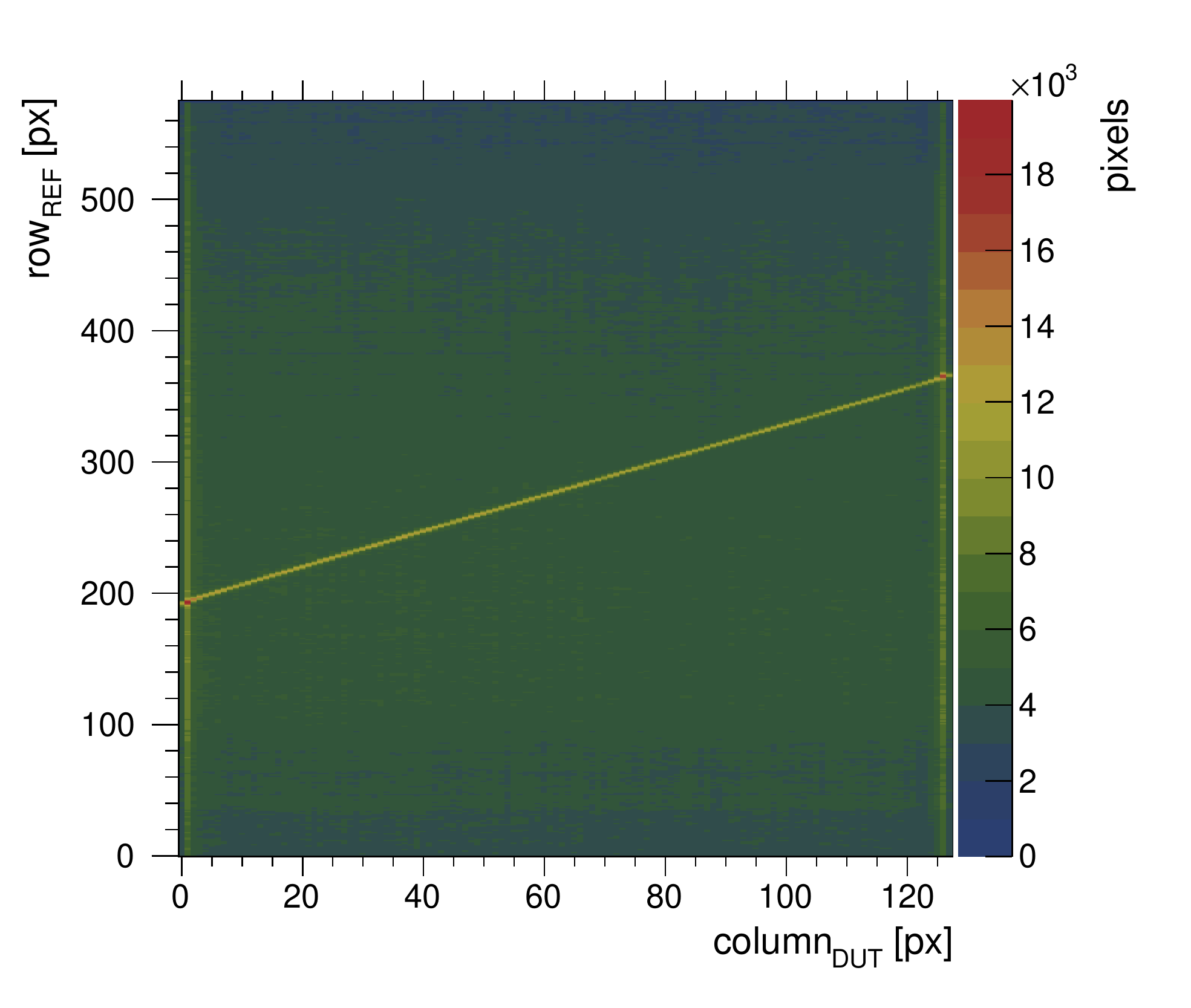}
      \put(20,68){\color{white}CLICdp}
    \end{overpic}
    \caption{Cross-correlations of the DUT $x$ axis}
    \label{fig:corr_x}
  \end{subfigure}%
  \begin{subfigure}[t]{0.5\linewidth}
    \begin{overpic}[width=\textwidth]{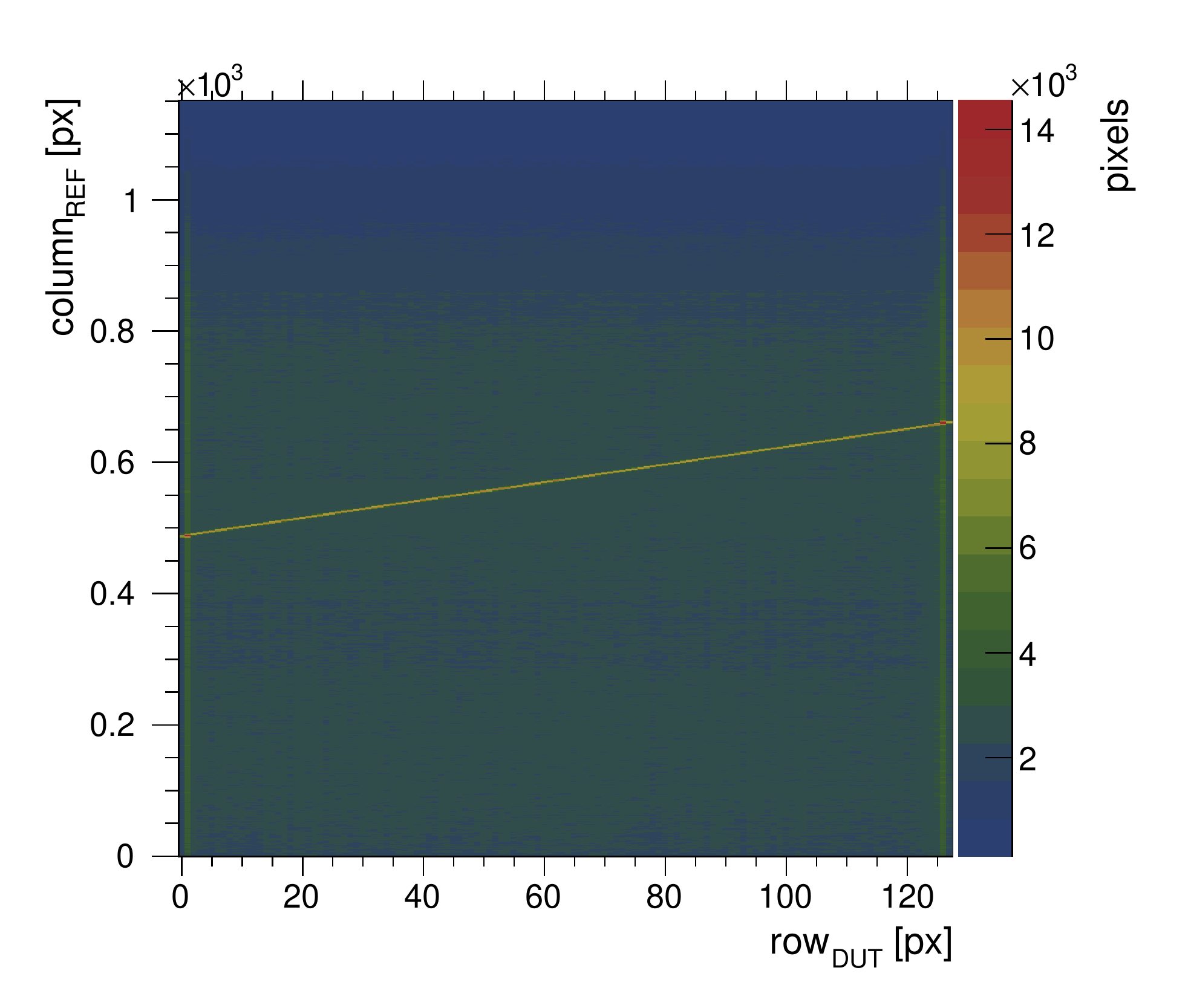}
      \put(20,68){\color{white}CLICdp}
    \end{overpic}
    \caption{Cross-correlations of the DUT $y$ axis}
    \label{fig:corr_y}
  \end{subfigure}%
  \caption{Cross-correlations between CLICpix2 (DUT) and the last upstream telescope plane (REF) in $x$~(\subref{fig:corr_x}) and $y$~(\subref{fig:corr_y}). The two detectors are rotated by \SI{90}{\degree} with respect to each other, which makes it necessary to compare opposite coordinates (columns versus rows). These plots are provided by the \emph{Correlations} module.}
  \label{fig:correlations}
\end{figure}

Correlation plots are an important means to gauge data quality, synchronisation between devices and mechanical alignment of the different detector planes.
Figure~\ref{fig:correlations} shows two cross-correlation plots produced by the \emph{Correlations} module.
Here, pixel hit positions are compared between the CLICpix2 DUT and the last upstream Mimosa26 plane of the beam telescope, which has been marked as reference detector in the \corry geometry description.
Shown are the hit column address of one detector versus the hit row address of the other detector and vice versa since the two planes are rotated by \SI{90}{\degree} with respect to each other.
A clear correlation is visible as bright line above a background of uncorrelated hits stemming from residual noise contributions as well as hits of the Mimosa26 plane outside of the DUT acceptance.
The line does not coincide with the diagonal of the histogram since the two detectors are different in size, with the CLICpix2 occupying an area of only \SI{3.2x3.2}{mm} in the centre of the \SI{20x10}{mm} large Mimosa26 sensors.

Projections of these correlation diagrams along the diagonal are used in the prealignment in order to correct the relative position of detectors with respect to the reference plane.

\paragraph{Tracking \& telescope alignment}

\begin{figure}[tbp]
  \centering
  \begin{subfigure}[t]{0.468\linewidth}
    \includegraphics[width=\textwidth]{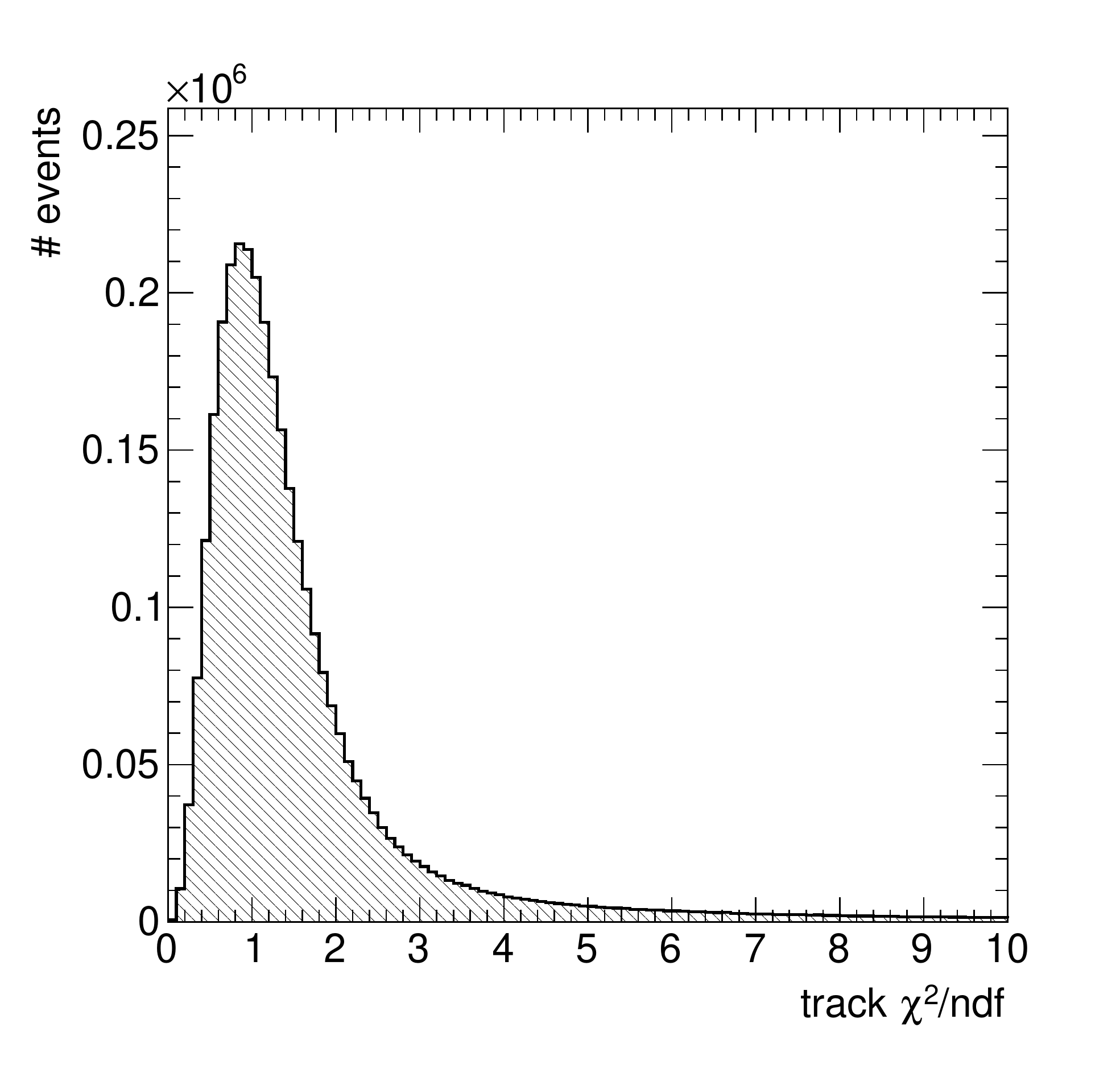}
    \caption{Track $\chi^2$ divided by the number of degrees of freedom, provided by the \emph{Tracking4D} module.}
    \label{fig:trackchi2}
  \end{subfigure}%
  \begin{subfigure}[t]{0.532\linewidth}
    \begin{overpic}[width=\textwidth]{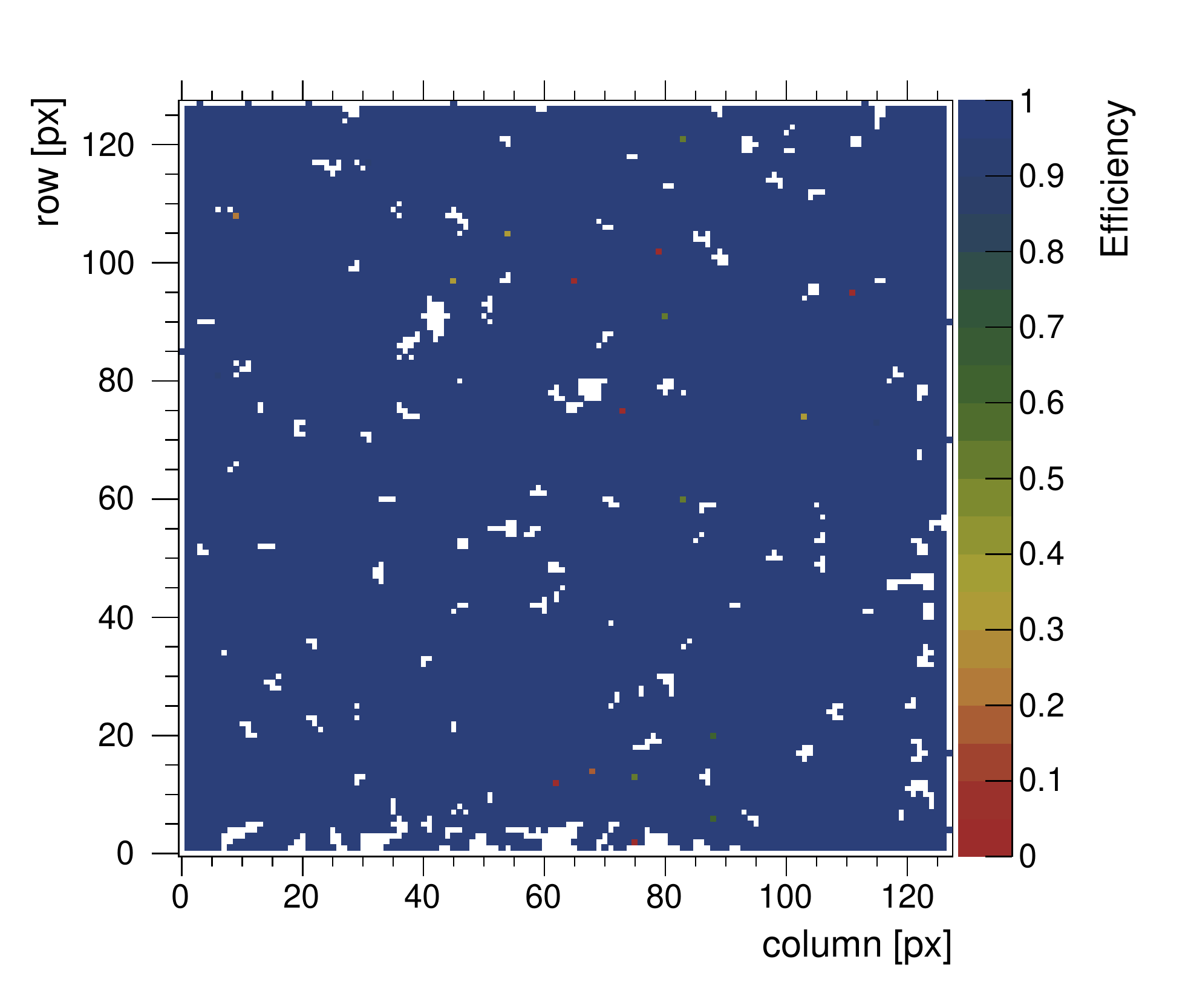}
      \put(20,68){\color{white}CLICdp}
    \end{overpic}
    \caption{Efficiency as a function of the track position over the full chip area, masked areas appear white.}
    \label{fig:efficiencymap}
  \end{subfigure}%
  \caption{General Broken Lines track $\chi^2$ as measure for the goodness of fit~(\subref{fig:trackchi2}) and CLICpix2 efficiency map~(\subref{fig:efficiencymap}).}
  \label{fig:tracking}
\end{figure}

Owing to the beam energy of \SI{5.4}{\GeV} used at the DESY~II Test Beam Facility and the expected multiple scattering along the particle trajectory, General Broken Lines has been selected as track model for the \emph{Tracking4D} module.
The module is configured to require hits from at least six out of seven detector planes available for tracking, while always requiring a hit from the Timepix3 time reference plane.
Subsequently, the cluster timestamp from the Timepix3 data is assigned as track timestamp.

The DATURA telescope planes and the Timepix3 detector are aligned using the iterative approach minimising the global $\chi^2$ value of the track fit implemented in the \emph{AlignmentTrackChi2} module.
Alignment is performed in the $x$ and $y$ position and all three orientation degrees of freedom.
The procedure of refitting tracks and optimising the detector placements and orientation is repeated until a sub-micron alignment precision is reached.
The $z$ position constitutes a weak mode and was fixed to the values measured in the beam area.
The $\chi^2$ distribution over the degrees of freedom resulting from the track fits with the final alignment is shown in Figure~\ref{fig:trackchi2} with a peak close to one, indicating a good fit of the track to the detector data.

\subsection{Analysis of the DUT Performance}

After tracking and alignment of the detectors of the reference beam telescope have been performed, the DUT data can be analysed in a separate run.
According to the modular structure of the framework, the analysis of the various performance parameters is also divided into individual modules.
This facilitates the flexible analysis of different prototypes and to adjust the configuration parameters.
In this example, the CLICpix2 data are first assigned to the tracks reconstructed from the reference beam telescope.
Afterwards, the efficiency and the spatial resolution are evaluated.
For the measurement of these observables, only telescope tracks with $\chi^2 / \mathrm{ndof} < 3$ have been selected within the respective analysis modules.

It should be noted that this section does not present a detailed analysis of the DUT but should serve only as an example for the capabilities and functionality of the \corry framework.
More detailed studies of the CLICpix2 prototype, using laboratory and test beam data, can be found elsewhere~\cite{morag_iprd, thesis_morag}.

\paragraph{DUT cluster association to tracks}

DUT clusters are assigned to telescope tracks using the \emph{DUTAssociation} module with the \emph{closest distance} matching criterion comparing the cluster edge to the track position.
The maximum matching distance is chosen as one pixel pitch in either direction, i.e. \SI{25}{\um}.

\paragraph{Efficiency}

\begin{figure}[tbp]
  \begin{subfigure}[t]{0.468\linewidth}
    \begin{overpic}[width=\textwidth]{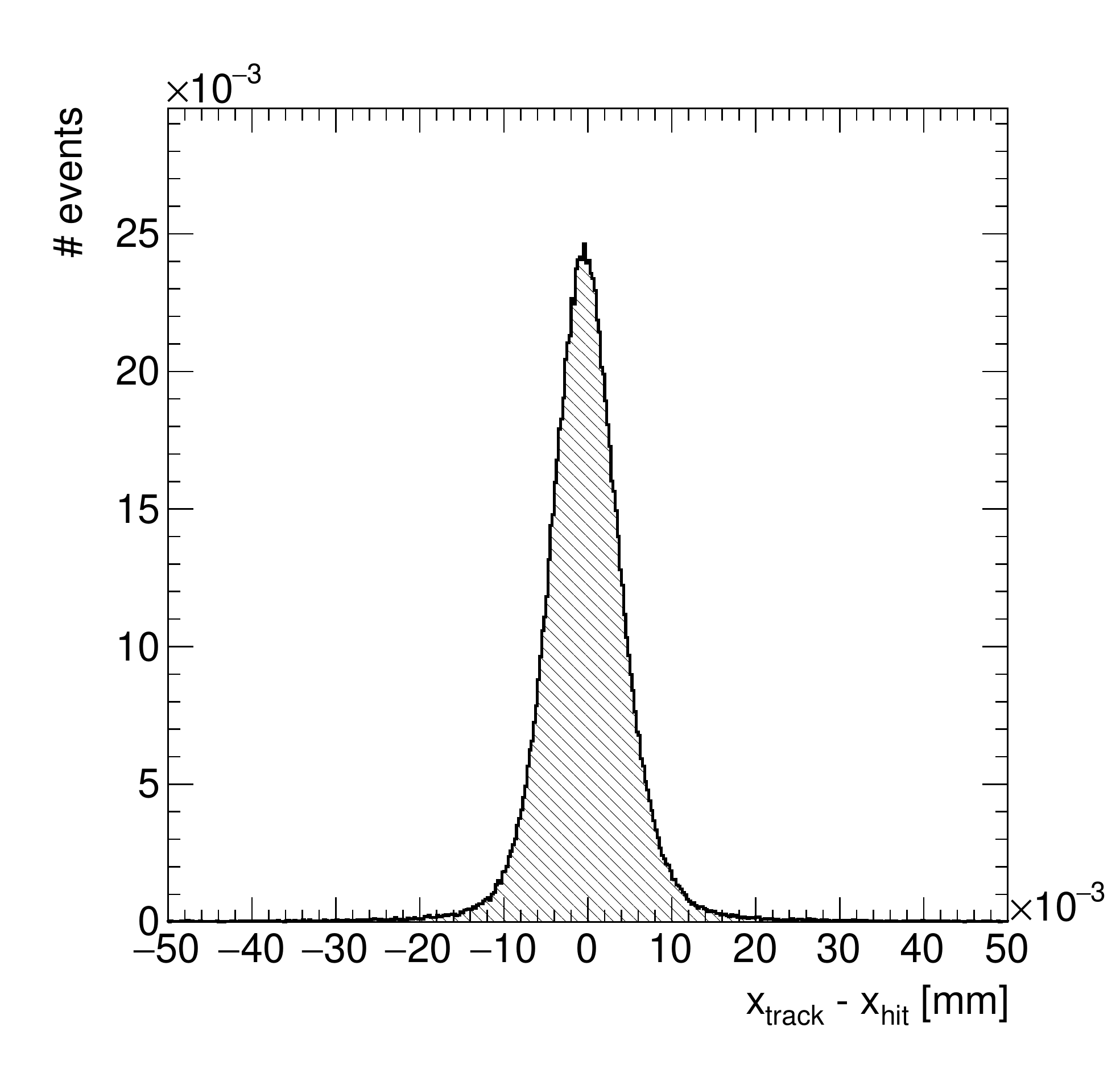}
      \put(20,78){CLICdp}
    \end{overpic}
    \caption{Unbiased residual distribution of the $x$ axis for CLICpix2, provided by the \emph{AnalysisDUT} module.}
    \label{fig:residual}
  \end{subfigure}%
  \begin{subfigure}[t]{0.532\linewidth}
    \begin{overpic}[width=\textwidth]{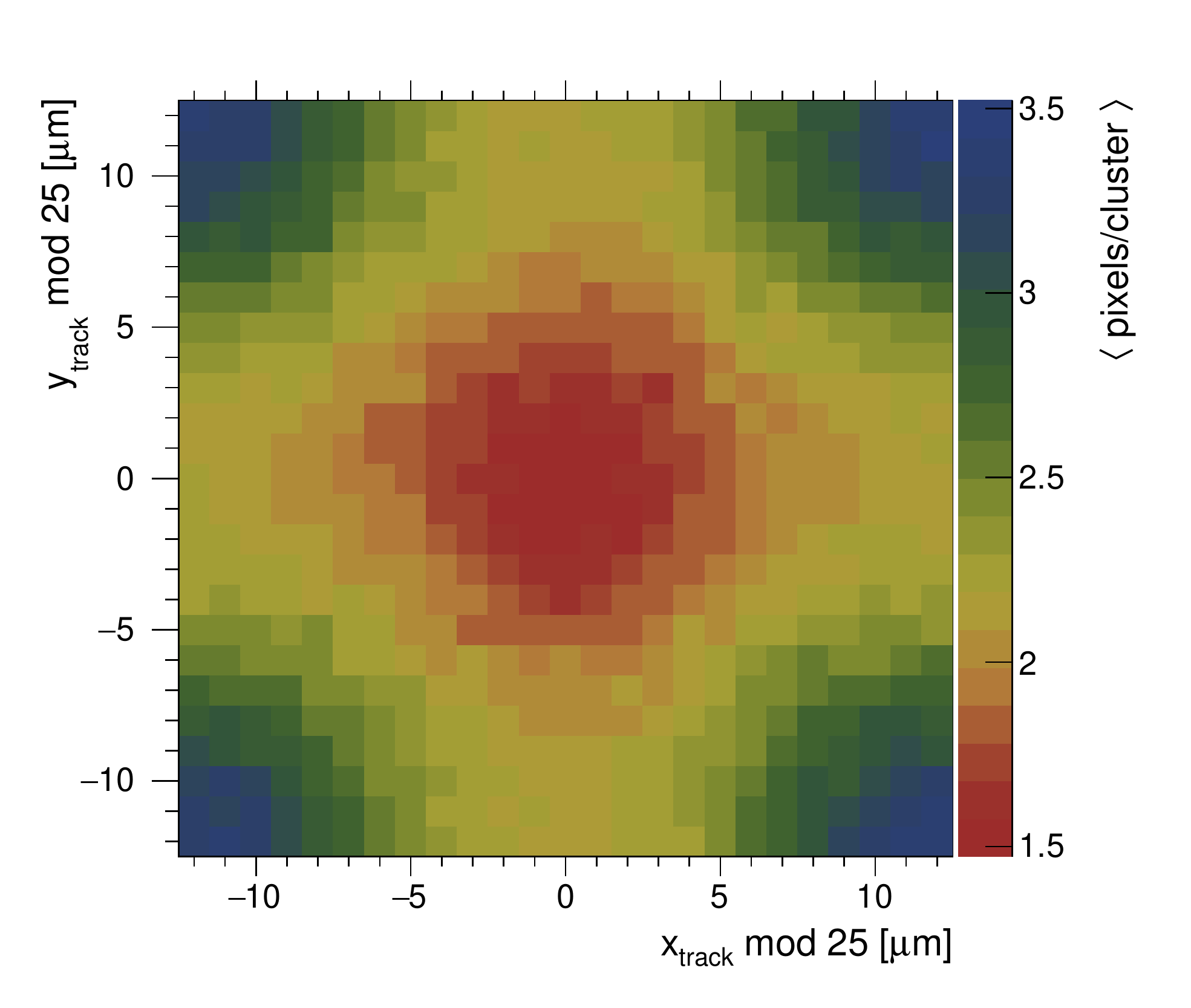}
      \put(20,68){\color{white}CLICdp}
    \end{overpic}
    \caption{Mean cluster size as function of the track impact position within a pixel cell.}
    \label{fig:npxvsxmym}
  \end{subfigure}%
  \caption{Unbiased residual distribution of CLICpix2 for clusters of all sizes~(\subref{fig:residual}) and mean cluster size~(\subref{fig:npxvsxmym}) at vertical track incidence within a single pixel cell of the CLICpix2 prototype.}
  \label{fig:efficiency}
\end{figure}

The efficiency of the DUT is calculated using the \emph{AnalysisEfficiency} module as the number of tracks with an associated DUT cluster divided by the total number of tracks passing through the detector.
Removed from this selection are tracks that penetrate the detector in areas with masked pixels with a one-pixel tolerance, as well as the outermost rows and columns of pixels close to the sensor edge.
Apart from these masked areas, the prototype exposes a very uniform efficiency as shown in Figure~\ref{fig:efficiencymap}, with only a few pixels that are less efficient.
The overall efficiency for the data analysed is evaluated to be above \SI{99.97}{\percent}.

\paragraph{Position resolution}

Finally, the position resolution of the detector can be determined via the \emph{AnalysisDUT} module by quadratically subtracting the telescope track resolution from the width of the residual distribution presented in Figure~\ref{fig:residual}.
Here, the residual width is defined as the root mean square of the truncated distribution containing the central \SI{96}{\percent} of the statistics, i.e.\ $\pm2\sigma$, and evaluates to $\sigma_{res} = \SI{4.1}{\um}$.

Using the high pointing resolution of the reference tracks, the \corry analysis modules facilitate the evaluation of different quantities as a function of the track incidence position within a single pixel cell of \SI{25x25}{\um} of the CLICpix2 DUT.
As an example, Figure~\ref{fig:npxvsxmym} show the average cluster size as a function of the track impact position within the pixel cell, which is a quantity used to gauge charge sharing between neighbouring pixel cells.
Single-pixel clusters occur almost exclusively in a clearly defined area in the centre of the pixel cell, since strong charge sharing effects arise from the ratio of pixel pitch and sensor thickness.
All tracks incident outside this area will form multi-pixel clusters by means of charge sharing.

%% file: conclusions.tex
In this paper, the test beam data reconstruction framework \corry has been presented.
It is a modular framework that enables the correlation of detectors with different readout schemes through its flexible event building algorithm.
Through a direct interface to the EUDAQ data acquisition system it is capable of handling any detector with the necessary conversion plugins available through the DAQ.

The reconstruction chain is built from individual modules, each performing a specific task or implementing a single algorithm.
A range of modules has been presented briefly, and the event building process has been described in detail using different application scenarios.
This includes setups with detectors using the same synchronisation method, like triggers or common timestamps, as well as a combination of these devices or advanced configurations using triggered shutters.
The joint operation of three different detectors and a Trigger Logic Unit was used as an example demonstrating the reconstruction and analysis of data recorded at the DESY~II Test Beam Facility.

Several extensions of the \corry framework as well as additional modules and track models are already under development in order to further extend the functionality and to serve an even wider community.
This includes features such as tracking in magnetic fields or the reconstruction of data from detectors with radial geometries, as well as parallel processing of data on multi-core machines.

%% file: acknowledgements.tex
This work was carried out in the framework of the CLICdp Collaboration.
The measurements leading to these results have been performed at the DESY~II Test Beam Facility at DESY Hamburg (Germany), a member of the Helmholtz Association (HGF).
This project has received funding from the European Union's Horizon 2020 Research and Innovation programme under Grant Agreement no. 654168.
We would like to thank Matthew Buckland, Manuel Colocci, Alexander Ferk, Adrian Fiergolski, Magnus Mager, Andreas N\"urnberg, Mateus Vicente Barreto Pinto and Jin Zhang who have provided valuable feedback and contributed to the \corry software.